\documentstyle[preprint,aps,eqsecnum,epsfig]{revtex}
\newcommand{\be}{\begin{equation}}
\newcommand{\ee}{\end{equation}}
\newcommand{\bea}{\begin{eqnarray}}
\newcommand{\eea}{\end{eqnarray}}
\tightenlines
\begin{document}
\preprint{PITT-99-??; LPTHE-99/40} 
\draft 
\title{\bf DYNAMICS OF SYMMETRY BREAKING IN FRW COSMOLOGIES: EMERGENCE
OF SCALING}  
\author{{\bf D. Boyanovsky$^{(a,b)}$ and  H. J. de Vega$^{(b,a)}$}}  
\address { (a)Department of Physics and Astronomy,
University of Pittsburgh, Pittsburgh, PA 15260 USA\\ (b)
LPTHE\footnote{Laboratoire Associ\'{e} au CNRS UMR 7589.}  Universit\'e Pierre
et Marie Curie (Paris VI) et Denis Diderot (Paris VII), Tour 16, 1er. \'etage,
4, Place Jussieu 75252 Paris, Cedex 05, France} 
\date{\today} 
\maketitle
\begin{abstract}
The dynamics of a symmetry breaking phase transition is studied in a
radiation and matter dominated spatially flat 
FRW cosmology in the large $ N $ limit of a scalar field theory. The
quantum density matrix is evolved from an initial state of quasiparticles in
thermal equilibrium at a temperature higher than the critical. The
cosmological expansion decreases the temperature and triggers the
phase transition. We identify  three different time scales: 
an early regime dominated by linear instabilities and the exponential
growth of long-wavelength fluctuations, an intermediate scale when the
field fluctuations probe the broken symmetry states and an asymptotic
 scale wherein a {\bf scaling regime} emerges for  modes of
wavelength comparable to or larger than the horizon. The scaling
regime is characterized by a dynamical physical correlation length 
$ \xi_{phys} = d_H(t) $ with $ d_H(t) $ the size of the causal horizon,
thus there is one correlated region per causal horizon. Inside these
correlated regions the field fluctuations sample the broken symmetry
states. The amplitude of the long-wavelength fluctuations becomes
non-perturbatively large due to  the early times instabilities 
and a semiclassical but {\em stochastic} description emerges in the 
asymptotic regime. In the scaling regime, 
the power spectrum is peaked at zero momentum revealing the onset of a
Bose-Einstein condensate. The scaling solution results in 
that the  equation of state of the scalar fields is the {\em same as
that of the background fluid}. This implies a Harrison-Zeldovich spectrum of 
scalar density perturbations for long-wavelengths.
We discuss the corrections to scaling  as well as
the universality of the scaling solution and the differences and
similarities with the classical non-linear sigma model.
\end{abstract}

\section{Introduction, Motivation and summary}

Symmetry breaking phase transitions play a fundamental role in the
Standard Model of particle physics and its generalizations and are 
conjectured to have ocurred in the early Universe. Some versions of
inflationary models invoke a symmetry breaking phase transition to 
generate the inflationary stage\cite{reviews} and the process of field
ordering after the phase transition has been argued to produce 
density fluctuations that could seed galaxy
formation\cite{reviews,kibble}. Certainly, symmetry breaking phase
transitions in the early Universe are 
very important at the electroweak scale and at much lower temperatures
for the hadronization and chiral phase  
transitions in QCD. At a more speculative level, late time phase
transitions in the matter dominated era had been
proposed\cite{latetimePT} with the possibility that pseudo
Nambu-Goldstone bosons may contribute to the  dark matter component
and influence large scale structure formation.  

Typically,  symmetry breaking phase transitions are studied by means of
an effective potential which is the 
free energy density for a space-time constant expectation value of the
order parameter. Although this formulation provides a good qualitative 
understanding of the {\em equilibrium} properties it is unable to
address the important issue of the {\em dynamics} of symmetry breaking 
and the non-equilibrium process of phase ordering and phase
separation. The simple picture of the order parameter 
rolling down the potential hill although intuitively appealing does
not capture the correct dynamics. A consistent 
non-equilibrium field theoretical treatment reveals that the quantum
fluctuations are extremely important and enhanced by 
the instabilities that are the hallmark of the phase
transition\cite{boyvega}.   

The picture that emerges from consistent studies of the dynamics of
phase transitions in Minkowski space  
time\cite{boyvega,symbreaklosala,bosecon,noscorre}   is remarkably
similar to phase ordering kinetics in a symmetry breaking 
phase transition in condensed matter systems\cite{nosreviews,bray,gill}. 

Recent results in quantum field theory\cite{bosecon,noscorre} in
Minkowski space-time reveal the following picture: 
when the system is cooled below the critical temperature on
 time scales shorter than the relaxation time of long-wavelength
fluctuations, the process of phase separation begins 
via linear spinodal instabilities and the exponential growth of these
long-wavelength fluctuations\cite{boyvega,bosecon,noscorre}. A new
dynamical time 
scale emerges at which the back-reaction of these fluctuations becomes
of the order of the tree level term and the 
linear instabilities shut-off. At this time scale, the fluctuations of
the field sample the broken symmetry equilibrium 
states (minima of the free energy) which in turn implies that the
amplitude of the long-wavelength fluctuations becomes 
non-perturbatively large\cite{boyvega,bosecon,noscorre} and consistent
non-perturbative methods are necessary to study the 
real-time dynamics of the phase transition. When the backreaction from
the field fluctuations becomes comparable to the 
tree level terms in the equations of motion the non-linearities become
very important. In this late time regime a dynamical 
length scale emerges $\xi(t)=t$ which is a consequence of causality
and determines the size of the correlated regions, 
the emergence of a scaling behavior and the onset of a novel
non-equilibrium condensate\cite{bosecon,noscorre,nosreviews,destri}. 
 Recently a very thorough study of
the non-equilibrium time evolution has been provided in finite volume
(finite momentum resolution) with non-perturbative 
techniques providing a consistent picture of the dynamics of phase
separation\cite{destri}.  

In cosmology the non-equilibrium dynamics of a symmetry breaking phase
transition has been studied within the 
context of {\bf inflationary} scenarios including quantum backreaction
from the matter fields in the metric dynamics,  
and non-equilibrium aspects of particle production and reheating
including quantum  backreaction in the large $ N $ limit had been
studied in Friedmann-Robertson-Walker (FRW) cosmologies in a fixed
background\cite{inflation,nuestros}. We investigate here the {\em  
dynamics of a phase transition and phase ordering kinetics} in a
radiation and in a matter dominated FRW background metric from a quantum
field theory formulation that includes self-consistently the quantum
and thermal fluctuations in the dynamics.  

Phase ordering dynamics  in a FRW background has been previously
studied in the large $ N $ limit of the {\em classical non-linear}
sigma model in\cite{turok,durrer}. Detailed numerical and analytical
studies of the classical non-linear sigma model revealed the emergence
of a scaling solution of the classical field equations that reflect
the dynamics of Goldstone bosons in the broken symmetry 
state\cite{turok}. The scaling solution allowed to study the cosmological 
perturbations seeded by the fluctuations of the Goldstone scaling
field\cite{turok,durrer} and provides a solvable example to study the
evolution of  causal cosmological perturbations.  
The importance of a scaling solution lies in the fact that for causal
perturbations the dynamical range of time scales 
between the phase transition and matter-radiation decoupling is very
large to be tractable numerically, the scaling property of correlation
function thus allows to extrapolate them to arbitrarily long time
scales\cite{kibble,durrer}.  

The {\em non-linear sigma model} is a theory that describes the
dynamics of Goldstone bosons in a broken symmetry state 
and as such the field is constrained to the vacuum manifold. Hence
this model {\em does not} allow to study the question of 
the dynamics of symmetry breaking phase transitions  and the
non-equilibrium aspects of phase separation and ordering 
 beginning from an initial high temperature state in the unbroken
symmetry phase and  cooling through the phase transition until
correlated regions of broken symmetry are formed and grow.  

{\bf The goals:} The goals and focus of this article are  to study the
important issue of the {\em dynamics of the phase transition} which
simply cannot be addressed within the context of the non-linear sigma
model. Namely, i) the 
non-equilibrium dynamics of a symmetry breaking phase transition from
an initial high temperature unbroken phase to 
a final low temperature broken symmetry phase in a radiation or matter
dominated background FRW cosmology, ii) to follow  the process of
phase separation and ordering directly in  real time from the early
time instabilities to the formation of correlated regions,  iii) to
study the emergence of a scaling regime as a dynamical consequence of 
the phase transition  and iv) implications for the equation of state
of the matter described by the quantum scalar field. Furthermore we analyze
the deviations from scaling and discuss the necessity for scaling
violations for self-consistency. The description of cosmological
perturbations  is outside the scope of this paper. We  keep our focus on the
description of the non-equilibrium dynamics of the phase transition and the 
resulting consequences. 

{\bf The model:} As argued above a consistent study of the dynamics of
a phase transition and the  process of phase separation requires a
non-perturbative framework. There are very few  non-perturbative
frameworks that are i) renormalizable ii) maintain all of the
conservation laws,  iii) lend themselves to a detailed analytical or numerical 
study and iv) can be consistently improved.  We thus study the {\em
quantum linear} sigma model of a scalar field in the 
vector representation of $O(N)$ in the leading order in the
large $ N $ limit in a fixed spatially flat radiation and or matter
dominated FRW cosmological background. There are several noteworthy
features of this model that deserve comparison with the classical
non-linear 
sigma model studied in detail in\cite{turok,durrer}: i) we consider
the full quantum field theory as described by a time 
dependent quantum density matrix prepared initially in equilibrium at
high temperature, ii) unlike the non-linear sigma 
model, this is a renormalizable quantum field theory and the
ultraviolet divergences are absorbed in renormalizations of
masses, couplings and field amplitudes, iii) the linear sigma model
allows to go from the unbroken 
to the broken symmetry phase, the non-linear model is only defined in
the broken symmetry phase.   

{\bf  The strategy:} The main ingredient in the description of the
non-equilibrium dynamics is the time evolution of 
an initially prepared quantum density matrix. We begin our analysis by
implementing the large $ N $ limit, which to leading order leads to a
self-consistent quadratic Hamiltonian with effective frequencies that 
depend on time through the background scale factor as well as through
the self-consistent mean field. The density matrix 
is taken to be Gaussian consistent with the leading order in the large
$ N $ and its initial form describes a state in which the
self-consistent quasiparticles are in local
thermal equilibrium at an initial temperature $ T_i>T_c $ with $ T_c $ the
critical temperature. We will not justify the 
choice  of an initial  thermal state for the self-consistent
quasiparticles, and we will simply {\em assume} that such initial state is 
physically reasonable and is an acceptable description of the system
prior to the phase transition. This initial density matrix evolves in
time through the Liouville equation which has an exact solution in the
leading order in the large $ N $ limit. The advantage of our
formulation in terms of the quantum density 
matrix is that it leads to a clear description of the emergence of 
(semi) classical but stochastic description.  
The expansion of the cosmological background results in adiabatic
cooling that triggers the phase transition resulting in long-wavelength 
instabilities which are the hallmark of the process of phase
separation and ordering and completely dominate the early time
evolution of correlations. The early time dynamics can be understood
analytically since for weak coupling the backreaction is
perturbatively small and the equations are essentially linear. The
intermediate and asymptotic dynamics are studied analytically and  
numerically both in radiation and matter dominated FRW backgrounds. 

{\bf Summary of results:} The main results of this article can be
summarized as follows:  

\begin{itemize}
\item{The dynamics of the phase transition reveals several different
time scales: an early time scale dominated by the exponential growth
of long-wavelength fluctuations as a consequence of the instabilities
associated with the phase transition. An intermediate time scale at which the
self-consistent backreaction from the quantum and thermal fluctuations
begins to compete with the tree-level 
terms in the equations of motion and signals the onset of
non-perturbative and non-linear dynamics. An asymptotic time scale and
the emergence of a scaling solution for fluctuations with wavelengths
of the order of or larger than the causal horizon. Corrections to
scaling are an unavoidable consequence of the dynamics and they are
important for self-consistency, their form is uniquely determined by
the background.}  

\item{In the scaling regime there emerges a dynamical length scale
$\xi_{phys}(t)=d_H(t)$ with $d_H(t)$ the size of 
the causal horizon. This length scale determines the size of the
correlated regions, inside which the field fluctuations 
probe the broken symmetry states. There is one such correlated region
per causal horizon. This is a microscopic, quantum 
field theoretical justification of Kibble's original
proposal\cite{kibble,kibble2}. The power spectrum is peaked at
superhorizon wavelengths and signals the onset of a non-equilibrium
condensate.  } 

\item{After the intermediate time regime the amplitude of
long-wavelength fluctuations become non-perturbatively large 
and the phases freeze out. A semiclassical but {\em stochastic}
description emerges and field configurations with 
large amplitudes and long-wavelengths are represented in the ensemble
(density matrix) with non-negligible probability. } 

\item{As a consequence of the scaling solution, the equation of state
for the fluid of the field fluctuations is the {\em same} as that of
the background. As a consequence, the density fluctuations $ \delta
\rho / \rho_{background} $ are time independent. Therefore, 
superhorizon models will exhibit a Harrison-Zeldovich spectrum.} 

\item{The universality of the scaling solution and the corrections
to scaling are discussed in detail. } 
\end{itemize}  

The article is organized as follows: In section II we set up the
model, implement the large $ N $ limit to leading order and 
solve the equations that determine the time evolution of the initially
prepared density matrix. In section III we study 
analytically and numerically the dynamics of symmetry breaking,
establishing the different time scales, the emergence of 
scaling and of a semiclassical description. In section IV the
consequences of the scaling solution are analyzed: the dynamical
length scale and the correlated regions and the equation of state of
the scalar fields fluctuations. Section V provides a comparison between
our results and those obtained for the {\em classical} $O(N)$
non-linear sigma model. Section 
VI summarizes our conclusions.

\section{Density Matrix evolution in the Large $ N $ limit}

We start by setting up the formulation  of the time evolution 
of the quantum density matrix in a spatially flat
FRW cosmology with a {\em fixed} 
background metric. In comoving coordinates the metric element is given by 
\begin{equation}
ds^2 = dt^2-a^2(t) \; d\vec{x}^2 \label{metric}
\end{equation}

with 

\bea
H^2(t)& = &  \left(\frac{\dot{a}}{a} \right)^2= \frac{8\pi}{3M^2_{Pl}}
\; \rho(t) \label{hubble} \\  
\rho(t) & = & \frac{\rho^i_R}{a^4(t)}+\frac{\rho^i_M}{a^3(t)} \label{rhoback}
\eea

\noindent and  $\rho(t)$ is a {\em fixed} background energy density
which is taken to be a combination of radiation and matter with
$\rho^i_R\; , \; \rho^i_M$ the energy densities in radiation and 
matter respectively at the initial time $t=t_0$ with $a(t_0)=1$. 

We consider a theory of $ N $ scalar fields  in the vector
representation of $O(N)$ in the leading order in the large $ N $ limit.
The action is given by 
\begin{eqnarray}
S         & =  & \int d^4x  \; a^3(t)\left\{\frac{1}{2}\dot{\vec
\Phi}^2(\vec{x},t)- 
\frac{1}{2 \; a(t)^2}[\vec{\nabla}\vec\Phi(\vec{x},t)]^2
-V(\vec \Phi(\vec{x},t))\right\}
 \label{lagrangian} \\
V(\Phi)   & =  & \frac{1}{2}[-m^2_0+ \xi_0 \; {\cal R}] \; {\vec \Phi}^2+
\frac{\lambda_0}{8N} \;(\vec \Phi \cdot \vec \Phi)^2
+\frac{m^4_0}{2\lambda}\label{potential} \\ 
{\cal{R}}    & =  & 6\left(\frac{\ddot{a}}{a}+\frac{\dot{a}^2}{a^2}\right)
\end{eqnarray}
with ${\cal{R}}$ the Ricci scalar and $\xi_0$ has been introduced with
the purpose of renormalization. Since the 
time evolution of the quantum density matrix is determined by
Liouville's equation, we need the Hamiltonian which 
requires the canonical momentum conjugate to $\Phi(\vec{x},t)$ 
\begin{equation}
{\vec \Pi}(\vec{x},t) = a^3(t)\; {\dot{\vec \Phi}}(\vec{x},t)
\label{canonicalmomentum} 
\end{equation}

The Hamiltonian becomes
\begin{equation}
H(t) = \int d^3x \left\{ \frac{{\vec \Pi^2(\vec{x},t)}}{2\;a^3(t)}+
\frac{a(t)}{2}(\vec{\nabla}\vec \Phi)^2+
a^3(t) \; V(\vec \Phi) \right\} \label{hamiltonian}
\end{equation}

In the Schr\"{o}dinger representation (at an arbitrary time $ t $),
the canonical momentum is represented as
$$ \Pi^a(\vec{x}) = -i \frac{\delta}{\delta \Phi^a(\vec{x})} \; \; ;
a=1,\cdots N  $$ 
and the functional density matrix  $\hat{\rho}$ with matrix elements in the
Schr\"{o}dinger representation 
$\rho[\Phi^a(\vec{.}), \tilde{\Phi}^b(\vec{.});t]$ obeys the Liouville
equation, which in the Schr\"{o}dinger representation 
becomes the functional differential equation\cite{FRW}

\be
i\frac{\partial }{\partial t}\rho[\Phi^a(\vec{.}),
\tilde{\Phi}^b(\vec{.});t] = \left( H\left[\frac{\delta}{\delta {\vec
\Phi}(\vec{.})}; \vec \Phi\right] -  H\left[\frac{\delta}{\delta
\tilde{\vec \Phi}(\vec{.})}; \tilde{\vec
\Phi}\right]\right)\rho[\Phi^a(\vec{.}), \tilde{\Phi}^b(\vec{.});t]
\label{liouville} 
\ee
To leading order the large $ N $ limit can be implemented by the
following Hartree-like factorization\cite{losalamos,nuestros} (for an
alternative formulation of the large $ N $ limit see\cite{losalamos}) 
\be
(\vec{\Phi}\cdot \vec{\Phi})^2 \rightarrow 2\, \langle \vec{\Phi}\cdot
\vec{\Phi}\rangle \; \vec{\Phi}\cdot \vec{\Phi}- \langle \vec{\Phi}\cdot
\vec{\Phi}\rangle^2  \; \; ; \; \; \langle \vec{\Phi}\cdot
\vec{\Phi}\rangle = N \langle \Phi^a \Phi^a \rangle \; (\mbox{no sum over}\; a)
\label{largenqft}
\ee
where the expectation value is in the time evolved density matrix
$\rho(t)$ which is the solution of the Liouville equation above. 

It is convenient to introduce the spatial Fourier transform of the fields as
\begin{equation}
\vec{\Phi}(\vec x,t) = \frac{1}{\sqrt{\Omega}} \sum_{\vec k}
\vec{\Phi}_{\vec k}(t) \;  e^{i \vec k \cdot \vec x} 
\end{equation}
with $\Omega$ the spatial volume, and a similar expansion for the
canonical momentum $\vec{\Pi}(\vec x,t)$. The Hamiltonian becomes
\begin{eqnarray}
H & = & \sum_{\vec k} \left\{ \frac{1}{2\;a^3(t)} \; \vec{\Pi}_{\vec
k}\cdot \vec{\Pi}_{-\vec k}+  
\frac{a^3(t)}{2}\; W^2_k(t)\; \vec{\Phi}_{\vec k}\cdot \vec{\Phi}_{-\vec k} - 
a^3(t)\; \frac{\lambda}{8N}\langle \vec{\Phi}\cdot \vec{\Phi}\rangle^2
+ \frac{m^4_0}{2\lambda} 
\right\} \label{hamqftlargen} \\ 
W^2_k(t) & = & -m^2_0+\xi_0 {\cal
R}+\frac{k^2}{a^2(t)}+\frac{\lambda}{2}\int \frac{d^3k}{(2\pi)^3}\;  
 \langle \Phi_{\vec k}^a \Phi_{-\vec k}^a  \rangle(t) ~~(\mbox{no sum
over} \; a)
\label{timefreqs} 
\end{eqnarray}
This Hamiltonian describes a set of infinitely many harmonic oscillators, that
are only coupled through the self-consistent 
condition in the frequencies (\ref{timefreqs}). Since the effective
Hamiltonian is quadratic in 
terms of the self-consistent frequencies, we propose  the following
Gaussian Ansatz for the functional density 
matrix elements in the Schr\"{o}dinger representation\cite{FRW}
\be
\rho[\Phi,\tilde{\Phi},t]  =  \prod_{\vec{k}} {\cal{N}}_k(t) \exp\left\{
- \frac{A_k(t)}{2} \; {\vec \Phi}_k \cdot {\vec \Phi}_{-k}-
\frac{A^*_k(t)}{2} \;\tilde{\vec \Phi}_k \cdot\tilde{\vec \Phi}_{-k}-
 B_k(t) \;{\vec \Phi}_k \cdot\tilde{\vec \Phi}_{-k} \right\} 
\label{densitymatrix} 
\ee
This form of the density matrix
is dictated by the hermiticity condition $\rho^{\dagger}[\vec
\Phi,\tilde{\vec \Phi},t] = 
\rho^*[\tilde{\vec \Phi},\vec \Phi,t]$; as a result of this condition
$ B_k(t) $ is real. 
The kernel $ B_k(t) $ determines the amount of mixing in the
density matrix. If $B_k=0$, the density matrix corresponds to a pure
state because it is a direct product of a state vector times its adjoint 
conjugate.  

The Liouville equation (\ref{liouville}) becomes\cite{FRW}
\begin{eqnarray}
& & i \frac{\partial }{\partial t}\rho[\vec \Phi,\tilde{\vec \Phi},t] =
\sum_{\vec k}\left\{ -\frac{1}{2a^3(t)}\left(
\frac{\delta^2}{\delta {\vec \Phi}_{\vec k} \cdot 
\delta {\vec \Phi}_{-\vec k}}-
\frac{\delta^2}{\delta \tilde{{\vec \Phi}}_{\vec k} \delta 
\tilde{{\vec \Phi}}_{-\vec k}}\right) \right.
 \nonumber \\
& & \left. +
\frac{a^3(t)}{2}\; W^2_k(t)\left({\vec \Phi}_{\vec{k}}\cdot 
{\vec \Phi}_{-\vec{k}}
-\tilde{{\vec \Phi}}_{\vec k} \cdot \tilde{{\vec \Phi}}_{-\vec k}\right)
\right\} \rho[\vec \Phi,\tilde{\vec \Phi},t] \label{liouvischroed}
\end{eqnarray}
The equations for the kernels in the density matrix ($ A_k, B_k $) are
obtained by comparing similar 
powers of $ {\vec \Phi}_{\vec k} $ on both sides of the above
equation. We obtain the following equations for the covariances:
\begin{eqnarray}
i\frac{\dot{{\cal{N}}}_k}{{\cal{N}}_k} & = & \frac{1}{2\, a^3(t)}(A_k-A^*_k)
\label{normeq} \\
i\dot{A}_k                     & = &
\frac{A^2_k-B^2_k}{a(t)^3}-a^3(t)\; W^2_k(t) \label{Aeq} \\
i \dot{B_k}                    & = & \frac{B_k}{a^3(t)}(A_k-A^*_k)
\label{Beq} 
\end{eqnarray}
The equation for $B_k(t)$ reflects the fact that an initial pure state with
$ B_k(0)=0 $ remains pure under time evolution. From these equations it
becomes clear that the only independent equation is that for
$A_k(t)$. In particular, 
writing $A_k(t)= A_{R,k}(t)+iA_{I,k}(t)$ we find that the ratios
$A_{R,k}(t)/B_k(t)$ 
and $ {\cal N}_k(t)/[A_{R,k}(t)+B_k(t)]^{\frac{1}{2}} $ are 
constant in time, the latter being a consequence of unitary time
evolution of the density matrix. Exploiting the 
proportionality between $ A_{R,k}(t) $ and $ B_k(t) $ we  introduce a
new variable\cite{FRW}
$$
{\cal A}_k(t) = {\cal A}_{R,k}(t) + i {\cal A}_{I,k}(t)
$$ 
by defining
\bea
A_{R,k}(t) &=& {\cal A}_{R,k}(t) \coth\Theta_k \label{areal} \\
B_{k}(t) &=& -\frac{{\cal A}_{R,k}(t)}{\sinh\Theta_k}
\label{bcoef}\\ 
A_{I,k}(t) &=& {\cal A}_{I,k}(t) \label{aimag}
\eea
This new variable obeys the following Ricatti equation of motion
\be
i{\dot{\cal A}}_k(t) = \frac{{\cal A}^2_k(t)}{a^3(t)} - a^3(t) \; W^2_k(t)
\label{Adot} 
\ee
This equation can be linearized by defining
\be
{\cal A}_k(t) = -i a^3(t) \; \frac{\dot{\varphi}^*_k(t)}{\varphi^*_k(t)}
\label{varphi} 
\ee
\noindent and the mode functions $\varphi_k(t)$ obey the following
equation of motion 
\be
\ddot{\varphi}_k(t) + 3 \; \frac{\dot{a}(t)}{a(t)}\;\dot{\varphi}_k(t)+
W^2_k(t)\;\varphi_k(t)=0 \label{modesk} 
\ee
The  equations of motion (\ref{modesk}) are recognized as the {\em
Heisenberg} equations of motion for $ \vec \Phi_{k}(t) $ 
obtained from the Hamiltonian (\ref{hamqftlargen}). This observation
leads to a rather clear interpretation of the mode 
functions $ \varphi_k(t) $ as a basis for the expansion of the
Heisenberg field operator $ \vec \Phi_{\vec k}(t) $, i.e. 
\be
\vec \Phi_{\vec k}(t) = \frac{1}{\sqrt{2}}\left[\; \vec{\bf a}_{\vec k}\;
\varphi_{ k}(t) + \vec{\bf a}^{\dagger}_{-\vec k}\;\varphi^*_k(t)\; \right]
\label{heisfield} 
\ee
where the annihilation and creation operators $\vec{\bf a}_{\vec k}~;~\vec{\bf a}^{\dagger}_{-\vec k}$ are independent
of time in the Heisenberg picture and define a Fock representation.
We will normalize the mode functions $\varphi_k(t)$ so that the
Wronskian is given by 
\be
{\cal W}[\varphi_k(t),\varphi^*_k(t)] =
\varphi^*_k(t)\dot{\varphi}_k(t) - \varphi_k(t)\dot{\varphi}^*_k(t) =
-2i  \left[\frac{ a(t_0)}{a(t)}\right]^3 \label{wronskian}
\ee
\noindent where $ t_0 $ is the initial time at which the density
matrix is prepared. Hence we find  
\be
 \langle \Phi_{\vec k}^a \;\Phi_{-\vec k}^a \rangle(t) =
\frac{|\varphi_k(t)|^2}{2a^3(t_0)}\coth\frac{\Theta_k}{2}\quad
,\quad \forall a=1,\ldots, N \label{expecval} 
\ee
Without loss of generality, we  choose $ a(t_0)=1 $ which can always
be done by a simple rescaling of lengths. Furthermore,
we will choose $ \Theta_k $ to reflect a thermal density matrix at the
time $ t_0 $ in terms of the self-consistent
frequencies which are discussed below. 

\subsection{Conformal time analysis, initial conditions and renormalization}

It is convenient to change variables to conformal time $\eta$ defined as
\be
\eta =  \int^t_{t_0} \frac{dt'}{a(t')}~~; ~~ \eta(t=t_0)=0 \; ,
\label{conftime}
\ee
which is chosen to vanish at the initial time.

The scale factor in conformal time $ C(\eta) $ and conformally
rescaled mode functions $ f_k(\eta) $ are given by  
\bea
&&C(\eta)  =  a(t(\eta))~~;~~ C(0)=1 \label{confscalefac} \\
&&f_k(\eta)  =  a(t)\; {\varphi_k(t) \over \sqrt{\omega_k}}\label{confmode} 
\eea
The conformally rescaled mode functions obey the Schr\"{o}dinger-like
differential equation 
\bea
&&\left[\frac{d^2}{d \eta^2}+k^2+C^2(\eta)\;M^2(\eta)\right]f_k(\eta) =
0  \label{confeqns}\\ 
&&M^2(\eta)= -m^2_0+ (\xi_0-\frac{1}{6})\;{\cal R}(\eta)+ 
\frac{\lambda}{8\pi^2} \int k^2 \; dk\;
\frac{|f_k(\eta)|^2}{C^2(\eta)}\, \omega_k
\,\coth\frac{\Theta_k}{2}
\label{effmass} \\ 
&& {\cal R}(\eta) = 6 \; \frac{C''(\eta)}{C^3(\eta)} \label{ricciconf}
\eea
\noindent where primes denote derivative with respect to conformal
time. We will choose  the following initial conditions on the mode functions
\bea
f_k(0)& = & \frac{1}{\omega_k} ~~; ~~ f'_k(0)= -i\label{inicond}\\ 
{\omega_k} & = & \sqrt{k^2+M^2(0)} \label{inifreqs} 
\eea
\noindent in this manner the mode functions at the initial time
represent positive frequency (particle) modes. We now choose
\be
\Theta_k = \frac{\omega_k}{T_i} \label{temperature}
\ee
so that the initial density matrix at $t=t_0 ; \eta=0$ describes a
statistical ensemble in local thermal equilibrium with 
 an initial temperature $T_i$ for the conformal modes
describing (quasi) particles of self-consistent frequencies
$\omega_k$. We will not try to justify this choice of initial state 
in local thermal equilibrium and simply assume that it provides a
physically reasonable description of the state prior to the phase transition.  

The renormalization aspects had already been studied in detail in
references\cite{FRW,inflation,baacke}  with the result that  the
quadratic and logarithmic divergences in terms of an ultraviolet
cutoff $ \Lambda $ can be absorbed in mass, coupling and conformal coupling
renormalization $ m^2_0 \rightarrow m^2_R;\lambda_0 
\rightarrow \lambda_R; \xi_0 \rightarrow \xi_R $. We refer the reader
to those references for details and highlight that the main result of
the renormalization program is that
\be
M^2(\eta)= 
-m^2_R+ (\xi_R-\frac{1}{6})\;{\cal R}(\eta)+ \frac{\lambda_R}{2} \langle
\psi^2 \rangle_R \label{renorcond} 
\ee
with 
\bea
&& \langle \psi^2 \rangle_R = I_R +J  \label{psi2} \\
&&I_R  = \int \frac{d^3k}{(2\pi)^3} \left\{
\frac{\omega_k  \mid f_k(\eta) \mid^2}{2\; C^2(\eta)} -
\frac{1}{2\;k\;C^2(\eta)} + \right. \nonumber \\
 &   & \left. \frac{\theta(k-K)}{4k^3}\left[
-\frac{{\cal{R}}}{6}- \left(\frac{C'(0)}{C(\eta)}\right)^2
- m^2_R+(\xi_R-\frac{1}{6}){\cal{R}}(\eta)+\frac{\lambda_R}{2}
\langle \psi^2 \rangle_R
\right] \right\}
\label{irren} \\
&&J   = \int \frac{d^3k}{(2\pi)^3}
\frac{\mid
f_k(\eta)\mid^2}{C^2(\eta)}\frac{\omega_k}{\exp{\frac{\omega_k}{T_i}}
- 1} 
\label{jota}
\eea
\noindent with $K$ an arbitrary renormalization scale\cite{FRW}.
 The explicit 
relation between bare and renormalized parameters can be found
in\cite{FRW,inflation,baacke}. The finite temperature contribution $J$
can be written by separating in the integral the large k behavior of the
mode functions $ f_k(\eta) \sim e^{-ik\eta}/{\sqrt{k}} $ 
for $ k >> C(\eta)\, M(\eta) $. Assuming that the initial temperature
$ T_i >> M(0) $ we can separate the contribution from the large 
wave-vectors $ k \geq T >> M(0) $ as in the hard-thermal loop
approximation in thermal field theory\cite{lebellac,parwani}. The high
temperature limit of $ J $ has been obtained in reference\cite{FRW},
the leading contribution is obtained in the  hard thermal loop limit  
\be
J_{HTL} = \frac{T^2_{eff}(\eta)}{12}~~;~~ T_{eff}(\eta) = T_i \;
 \frac{C(0)}{C(\eta)} ~~;~~ T_{eff}(0)=T_i \label{HTL} 
\ee

This term is completely determined the very short wavelength modes $k
\geq T \gg M(0)$ and reflects the adiabatic {\em cooling} of 
the short wavelength modes in the initial state by the cosmological expansion. 
 
Introducing the critical temperature as
\be
T_c = m_R \sqrt{\frac{24}{\lambda_R}} \label{Tc} 
\ee 
we find that the hard thermal loop contribution $J_{HTL}$ combines
with the renormalized mass term $-m^2_R$ in $M^2(\eta)$
(\ref{effmass}) to yield  
\be
-m^2_R+\frac{\lambda_R}{2}J_{HTL}(\eta) = m^2_R\left
[ \frac{T^2_{eff}(\eta)}{T^2_c}-1 \right] \label{TminusTc} 
\ee
Clearly the leading order finite temperature contribution, i.e. the
hard thermal loop limit gives rise to an effective $\eta$ dependent 
mass squared term that changes sign when the effective temperature
falls below the critical, thus triggering the phase transition.  

There are sub-leading corrections in the high temperature limit that have been 
studied in detail in reference\cite{FRW}. The term linear in the
initial temperature corresponds to the {\em classical} contribution
for $\omega_k \ll T_i$, i.e. 
$k \ll T_i$ which is given by 
\be 
J_{cl}(\eta) = T_i \int \frac{d^3k}{(2\pi)^3}
\frac{\mid f_k(\eta)\mid^2}{C^2(\eta)} \label{jotaclassical}
\ee
The contributions from $ J_{HTL}(\eta) $ and $ J_{cl}(\eta) $ are the most
important ones in the weak coupling limit $\lambda_R \ll 1$ as 
can be seen as follows. The leading order correction  $J_{HTL}(\eta)$
determines the effective squared mass, which is positive for 
$T_{eff}(\eta) > T_c$ and becomes negative for $T_{eff}(\eta)<T_c$
thus triggering the phase transition. Once the effective temperature 
falls below the critical, long-wavelength instabilities begin growing
exponentially and the long-wavelength modes acquire non-perturbatively 
large amplitudes as argued in the introduction and studied explicitly
in detail below. These long-wavelength modes with $k<<T_i$  
determine the contribution $J_{cl}(\eta)$, which is the dominant
contribution from the infrared sector in the high temperature limit.  
In terms of the critical temperature (\ref{Tc}) and  the ratio
$T_i/T_c \geq 1$ the classical contribution leads to 
\be
\frac{\lambda_R}{2} J_{cl}(\eta) =  m_R\;
\frac{\sqrt{{24}{\lambda_R}}}{4\pi^2}\; \frac{T_i}{T_c}\;
 \int^{\kappa}_0 k^2 \;dk \; \frac{\mid f_k(\eta)\mid^2}{C^2(\eta)}
\label{jotacutoff} 
\ee
\noindent where $\kappa$ is a cutoff that determines the maximum
wavevector that will be unstable. Numerically it is found that the 
integral becomes insensitive to the choice of this cutoff for $\kappa
\geq m_R$. The next contribution is logarithmic in
temperature\cite{FRW} and is  of order $\lambda_R \ln(\lambda_R)$,
therefore subleading 
in the high temperature and weak coupling limit. Furthermore, the zero
temperature part is finite after renormalization and subleading 
in the weak coupling limit since it is of ${\cal O}(\lambda_R)$ as
compared to the ${\cal O}(1);{\cal O}(\sqrt{\lambda_R})$ for the 
${\cal O}(T^2_i); {\cal O}(T_i)$ contributions from $J_{HTL}; J_{cl}$
respectively. Hence the zero temperature part that remains after
renormalization will be neglected. 

We finally obtain the equations of motion in leading order in the high
temperature and weak coupling expansion after the following steps 

\begin{itemize}
\item{ Rescaling all dimensionful quantities, $k,\eta, f_k(\eta),
\omega_k$, etc. in terms of the only dimensionful parameter
$m_R$. Effectively this amounts to setting $m_R=1$ and all
dimensionful parameters are understood in units of $m_R$.}  

\item{ After renormalization of couplings and mass we neglect the zero
temperature contribution, which is higher order in the 
weak coupling $\lambda_R$  and keep only the leading order terms
$J_{HTL}(\eta)$ and $J_{cl}(\eta)$ neglecting higher order
contributions which are subleading in the high initial temperature and
weak coupling limits.}

\item{ We choose the case of minimal coupling $\xi_R=0$.} 

\end{itemize}
In summary, the equations of motion for the mode functions and 
the initial conditions are
\bea
&&\left[\frac{d^2}{d \eta^2}+q^2+C^2(\eta)M^2(\eta)\right]f_q(\eta) =
0  \label{finconfeqns}\\ 
&&M^2(\eta)=  \frac{T^2_{eff}(\eta)}{T^2_c}-1
-\frac{C''(\eta)}{C^3(\eta)}+g\Sigma(\eta) \label{massofeta} \\ 
&& \Sigma(\eta) =
 \int^{1}_0 {q^2 dq}\; \frac{\mid f_q(\eta)\mid^2}{C^2(\eta)} ~~; ~~
g=\frac{\sqrt{{24}{\lambda_R}}}{4\pi^2}\; \frac{T_i}{T_c}\;
 \label{gsigma}\\
&&f_q(0)= \frac{1}{\omega_q} ~~; ~~ f'_q(0) = -i \omega_q\; f_q(0)
\label{inicondfin}\\ 
&&\omega_q= \sqrt{q^2+\frac{T^2_i}{T^2_c}-1} \label{omegafin}
\eea 
where we have set the cutoff $\kappa = m_R \equiv 1$ which will be
justified numerically below, and neglected perturbatively small
corrections of ${\cal O}(\sqrt{\lambda_R})$ in $\omega_q$. The term
$g\Sigma(\eta)$ describes the back-reaction of the 
quantum and thermal fluctuations. 

With the background energy density determined by 
the combination of matter and radiation as given in
eq. (\ref{rhoback}) a simple expression is obtained for $C(\eta)$ by
introducing the quantities 
\be
r_i= \frac{\rho^i_M}{\rho^i_R} ~~; ~~ \frac{H^2_i}{1+r_i} = \frac{8\pi
\rho^i_R}{3M^2_{Pl}} \label{rHi} 
\ee
where we have used $a(t_0)=1$. A straightforward integration leads to
\be
C(\eta) = 1+ H_i\; \eta + \frac{H^2_i~ r_i}{4(1+r_i)}\;\eta^2 = \left\{ 
\begin{array}{cc} 
 1+ H_i\; \eta & ~\mbox{for Radiation dominated} \\
 1+ H_i\; \eta+ \frac{H^2_i}{4} \; \eta^2 &  ~\mbox{for Matter dominated} 
\end{array} \right. \label{Cofeta}
\ee
Rescaling $H_i$ and $\eta$ in terms of $m_R$ we choose the value
$H_i=1/2$ for  convenience. This value is motivated by a choice of an
initial radiation energy density $\sim (10^{16})^4~ (\mbox{Gev})^4$
with a mass scale $ \sim 10^{12}\mbox{Gev} $. Other values of $H_i$
lead to a proper re-scaling of the time scales and therefore a  
quantitative change in the results, but the qualitative features
described below are robust.  

\section{Dynamics of Symmetry Breaking:}
We begin our study of the dynamics of symmetry breaking by focusing on
the case of a radiation dominated FRW cosmology, 
i.e. we set $r=0$ in (\ref{Cofeta}) and as explained above we have
chosen $H_i =1/2$ for convenience. The dynamics is completely
determined by the set of equations (\ref{finconfeqns}-\ref{omegafin})
with $C(\eta)= 1+ \eta/2 $.  

Before embarking on a numerical study of these equations we can obtain
insight into the dynamics by neglecting the 
backreaction term $ g\Sigma(\eta) $ in (\ref{massofeta}). This
approximation is valid in the weak coupling limit and 
for early times when $ g\Sigma(\eta)\ll 1 $. During the time
scale for which $T_{eff}(\eta) > T_c$ the 
effective squared mass is positive and the mode functions are
oscillatory functions of conformal time. When the 
effective temperature falls below the critical, the effective squared
mass becomes negative and long-wavelength modes with momentum
$q^2 < C^2(\eta)\;M^2(\eta)$ become unstable and begin to grow almost
exponentially, obviously the growth rate is the 
largest for the smaller $q$. We can obtain an estimate of the behavior
of the long-wavelength modes after the temperature 
falls below the critical via a WKB expansion which after approximating
$M^2(\eta) \approx -1-C''(\eta)/C^3(\eta)$ leads to  
\be
f_{q\sim 0}(\eta) \sim {e^{\int^{\eta}C(\eta')\,|M(\eta')|\,d\eta'}
\over \sqrt{C(\eta)\,|M(\eta)|}} \label{WKB}
\ee 
\noindent which reflect an exponential growth of long-wavelength
fluctuations as a result of {\em linear instabilities} of the
long-wavelength modes below the critical temperature.  
This behavior in turn entails that the backreaction term also grows
exponentially as 
\be
g\Sigma(\eta) \sim g \frac{e^{2\int^{\eta}C(\eta')\,|M(\eta')|\,d\eta'}
}{C^3(\eta)\,|M(\eta)|} \label{gsigmalineal} 
\ee
and becomes non-perturbatively large at a time scale $\eta_{NL}$ when
it begins to compete with the tree level term, i.e. when  
\be
g\Sigma(\eta_{NL}) \sim 1 \label{endoflineal}
\ee
The condition (\ref{endoflineal}) 
determines the time scale $\eta_{NL}$ at 
which the non-linearities become important and must be treated
non-perturbatively. This condition has a simple 
physical interpretation: it implies that the amplitude of the long
wavelength fluctuations probe the broken symmetry states, i.e. since 
$$ 
g\Sigma(\eta) \approx \frac{\lambda_R}{2} \langle \Phi^a \Phi^a \rangle_R 
$$
restoring $ m_R $ as the scale unit we see that
$$
g\Sigma(\eta_{NL}) \approx m^2_R \Rightarrow \langle \Phi^a \Phi^a
\rangle_R \approx 2 \; m^2_R/\lambda_R \sim \Phi^2_0
$$  
with $ \Phi^2_0 $ being the minimum of the renormalized tree level
potential. Once $g\Sigma \approx m^2_R$ the backreaction compensates for
the negative squared mass and the long-wavelength 
instabilities shut-off. This picture is confirmed by a detailed
numerical analysis of the equations of motion  
(\ref{finconfeqns})-(\ref{omegafin}) with $ C(\eta)= 1+ \eta/2 $ for a
radiation dominated FRW cosmology $(r=0)$.   

\begin{figure}
{\epsfig{file=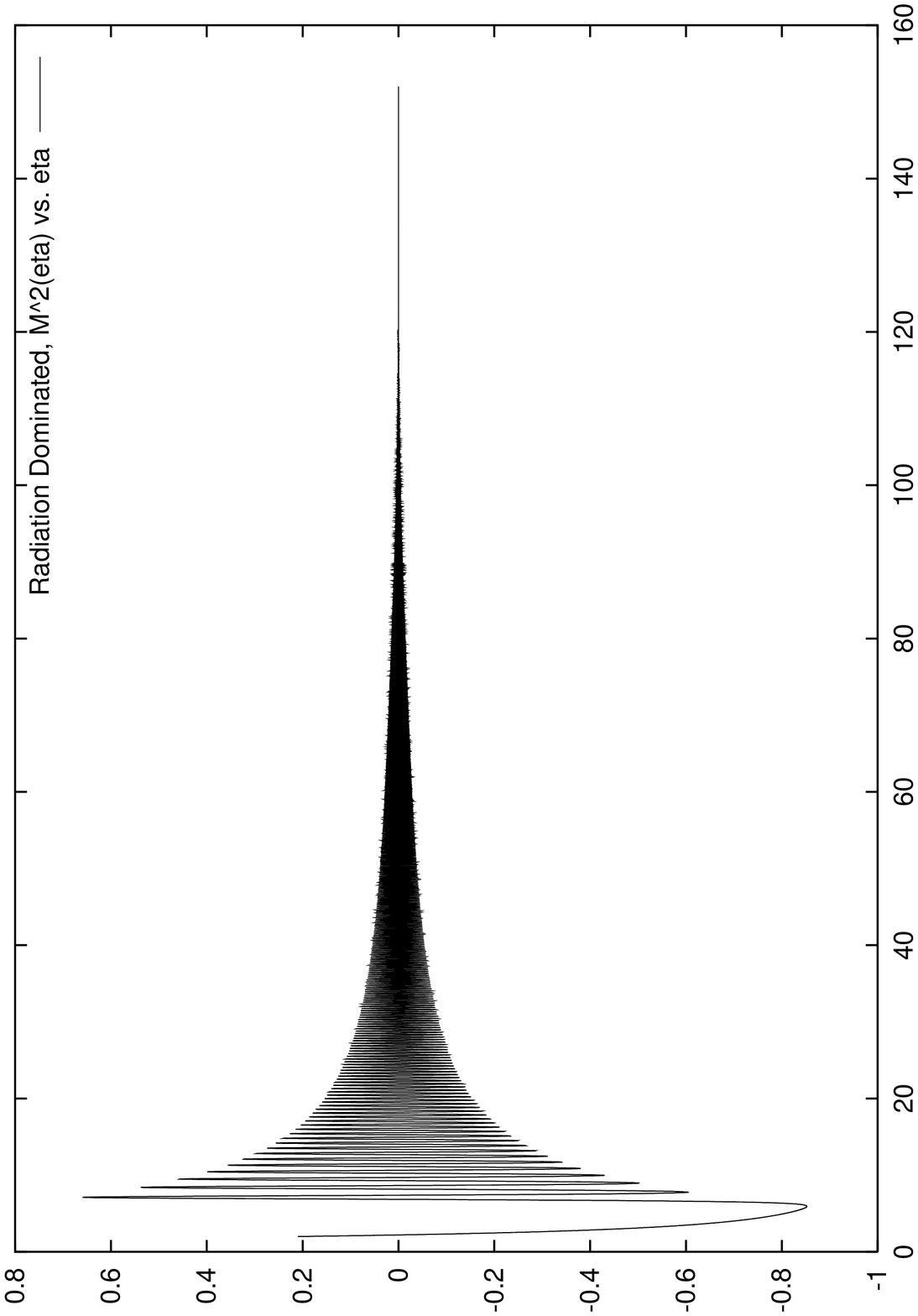,width=18cm,height=20cm}}
\caption{ $ M^2(\eta)$ vs. $ \eta $ (conformal time in units of
$ m^{-1}_R $) for $ \frac{T_i}{T_c}=1.1 $, $ g=10^{-5} $. Radiation
dominated universe. \label{fig1} } 
\end{figure}

\begin{figure}
{\epsfig{file=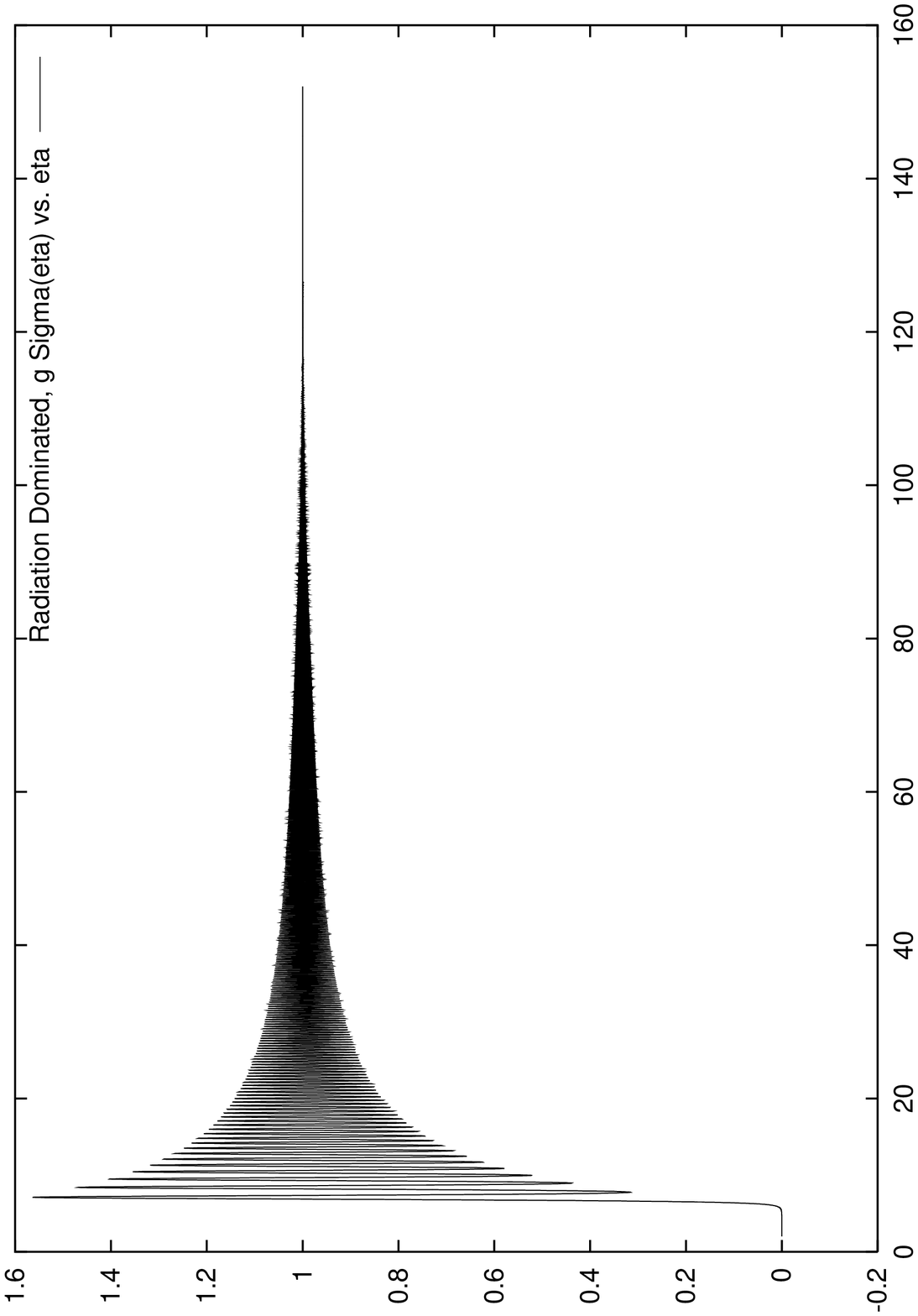,width=17cm,height=20cm}}
\caption{ $g\Sigma(\eta)$ vs. $\eta$ (conformal time in units of
$m^{-1}_R$) for $\frac{T_i}{T_c}=3$, $g=10^{-5}$. Radiation dominated universe.
\label{fig2} } 
\end{figure}

\begin{figure}
{\epsfig{file=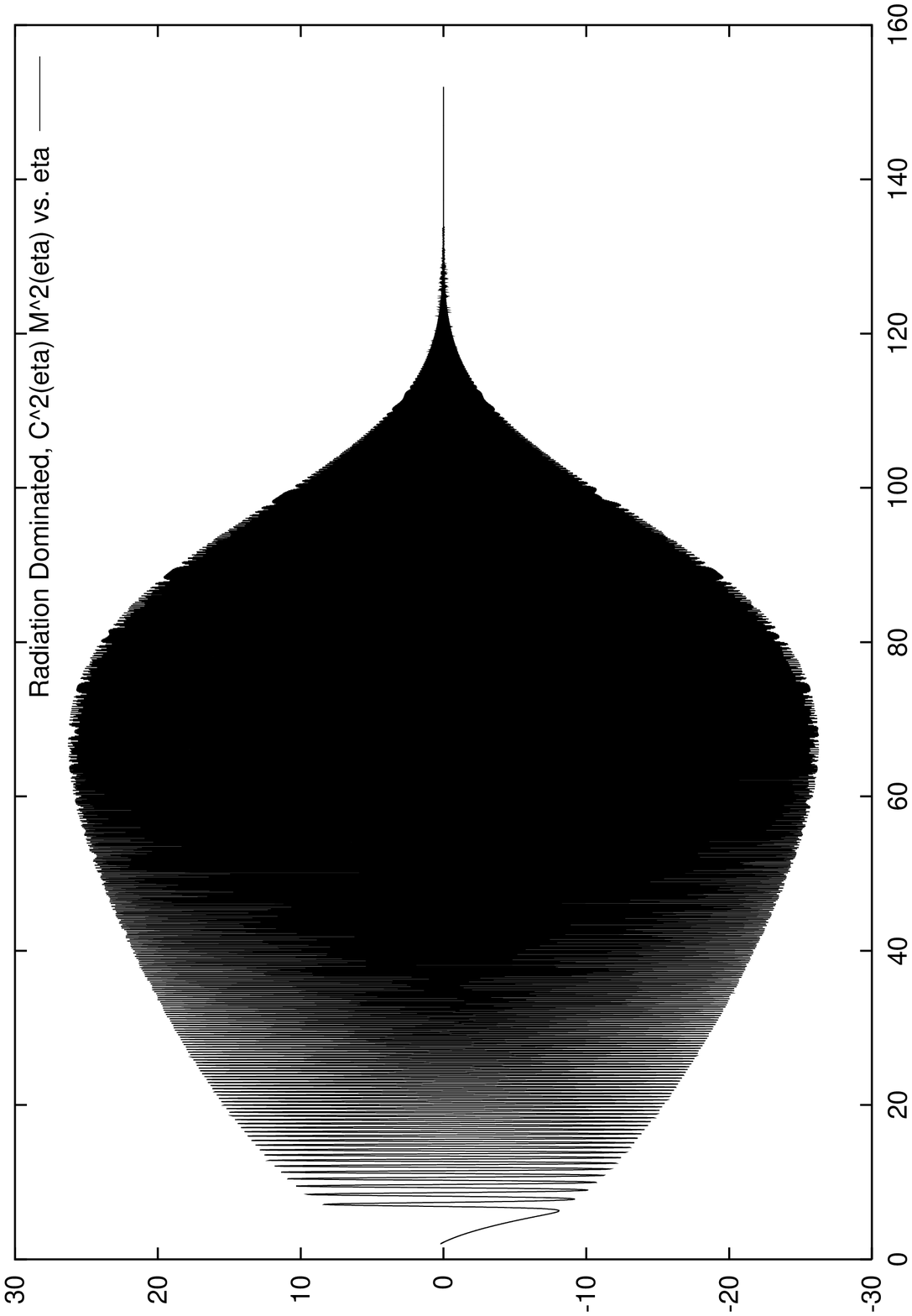,width=17cm,height=20cm}}
\caption{ $ C^2(\eta) M^2(\eta) $ vs. $ \eta $ (conformal time in units of
$ m^{-1}_R $) for $ \frac{T_i}{T_c}=1.1 \, , g=10^{-5}$. Radiation
dominated universe. \label{fig3} } 
\end{figure}

Fig. \ref{fig1} displays the effective squared mass $M^2(\eta)$ given
by eq. (\ref{massofeta}) for a 
radiation dominated ($r=0$) cosmology with the value of the parameters
$ T_i/T_c=3~;~g=10^{-5}~;~H_i=1/2 $ and all dimensionful quantities 
had been rescaled by $m_R$. The effective squared mass begins with
a positive value at $\eta=0$ since $T_i>T_c$  but diminishes as a
consequence of the adiabatic cooling through the cosmological
expansion. 
The back reaction $g\Sigma(\eta)$ is displayed in Fig. \ref{fig2}, it
is initially perturbatively small during the stage in  
which $ T_{eff}(\eta) > T_c $.  Once the effective temperature falls
below the critical, long-wavelength fluctuations begin to grow  
approximately as in eq.(\ref{WKB}) as a result of the linear instabilities
and the back reaction term begins 
to grow exponentially and to compete with the tree level term. The
rapid exponential growth of the back-reaction terms results in 
an overshooting and the effective mass $M^2(\eta)$ becomes positive.
When this happens the mode functions become again oscillatory 
and their amplitude diminishes causing a damped oscillatory behavior
in $ g\Sigma(\eta) $ and consequently in $ M^2(\eta) $. However, in
the Schr\"{o}dinger-like equations for the mode functions
(\ref{finconfeqns}) the mass term is $ C^2(\eta)M^2(\eta) $, thus while
$ M^2(\eta) $ becomes positive and oscillatory and eventually damps out
by dephasing of the oscillations the prefactor $C^2(\eta)$ makes 
this effective mass to grow during some period of time. The effective
mass $ C^2(\eta)M^2(\eta) $ for the mode functions $ f_q(\eta) $ is 
displayed in Fig. \ref{fig3} which clearly shows the initial cooling,
the growth of the backreaction and the competition between 
the damping of the oscillations in $M^2(\eta)$ and the growth of the
scale factor. This figure also reveals the striking asymptotic 
behavior that $C^2(\eta) M^2(\eta)\rightarrow 0$ as $\eta \rightarrow
\infty$. A detailed numerical analysis reveals that asymptotically 
at long time $C^2(\eta)M^2(\eta) \rightarrow {\cal O}(1/\eta^2)$, a
result that will be understood analytically below. The combination 
of figures (\ref{fig1}-\ref{fig3}) reveals three different time scales
after the phase transition, i.e. when $ T_{eff}<T_c $

\begin{itemize}
\item{Early time ($\eta < \eta_{NL}$): this stage is dominated by
linear instabilities that result in the exponential growth of
long-wavelength modes. The back-reaction contribution $ g\Sigma(\eta)
$ grows exponentially but remains perturbatively small.}  

\item{Intermediate time ($\eta \sim \eta_{NL}$): during this stage the
back-reaction begins to 
be comparable to the tree level contribution to the mass, i.e. 
$ g\Sigma(\eta) \sim m^2_R $, the back-reaction begins to shut-off the
linear instabilities but overshoots resulting in an oscillatory 
effective mass. This stage determines the onset of non-linear
evolution since the back-reaction is of the same order as the 
tree level term. The damping of $ M^2(\tau) $ competes with the growth
of the scale factor $ C^2(\eta) $ resulting in that the effective
squared mass $ C^2(\eta)M^2(\eta) $ still grows. }

\item{Asymptotic regime ($\eta \gg \eta_{NL}$): This regime is fully
non-linear and the detailed numerical analysis reveals that the effective mass
$ C^2(\eta)M^2(\eta) $ {\em vanishes} asymptotically. Numerically we
find that in this regime 
\be \label{asireg}
C^2(\eta)M^2(\eta) \buildrel{\eta \gg 1 }\over= -{(2p+3)(2p+1) \over 4
\eta^2} ~~~ \mbox{for} \; C(\eta) = \left(\frac{\eta}2\right)^p
\ee
\noindent with $p=1,2$ for radiation and matter dominated
respectively. The result (\ref{asireg}) implies a very delicate
mechanism of cancellation between the back-reaction and tree level
terms, which will be understood analytically below.  }   

\end{itemize}

\subsection{Asymptotic dynamics and emergence of scaling:}
From the expression for $ M^2(\eta) $ given by eq. (\ref{massofeta})
the vanishing of $ C^2(\eta)M^2(\eta) $ in the asymptotic region, 
when $T_{eff}(\eta)\ll T_c$ leads to the following sum rule in the
asymptotic regime $ \eta > \eta_{NL} \gg 1 $ (in units of $ m_R $) 
\be
g\int^{\infty}_0 {q^2 \; dq} \mid f_q(\eta)\mid^2 = C^2(\eta) \buildrel{\eta
\gg 1 }\over=  \left(\frac{\eta}2\right)^{2p}\label{sumrule} 
\ee
where $ p = 1 $ for radiation dominated and $ p = 2 $ for matter dominated.

The constraint that the integral on the left hand side must lead to a
power law suggests the following {\em scaling ansatz} for the mode functions
\be
|f_q(\eta)|^2 = \eta^{\alpha} \; |{\cal F}(x)|^2 ~~;~~ x = q\eta
\label{scaling} 
\ee
Comparing the powers and the coefficients leads to the following constraints
\be\label{power} 
\alpha = 2 p + 3
\ee
\be \label{coeff} 
g\int^{\infty}_0  x^2 \; dx \; |{\cal F}(x)|^2 = 2^{-2p}
\ee
Assuming that the scaling function ${\cal F}(x)$ is regular at $x=0$,
the scaling {\em ansatz} (\ref{scaling}) leads to a remarkable 
conclusion: the $q=0$ mode function is of the form
\be
f_0(\eta) = \eta^{\frac{\alpha}{2}}{\cal F}(0) \label{modocero}
\ee
leading to the result
\be
C^2(\eta)M^2(\eta) = - \frac{f''_0(\eta)}{f_0(\eta)} = -
\frac{(2p+3)(2p+1)}{4\eta^2} = \left\{ \begin{array}{cc} 
 -15/4\eta^2 & ~\mbox{for Radiation dominated} \\
-35/4\eta^2 &  ~\mbox{for Matter dominated} 
\end{array} 
\right.  \label{masaasyntotica} 
\ee
which is precisely the numerical result (\ref{asireg}).
The result (\ref{masaasyntotica}) in turn leads to the following
equations of motion for the $ q\neq 0 $ mode functions
\be
\left[\frac{d^2}{d\eta^2}+ q^2 +
\frac{(2p+3)(2p+1)}{4\eta^2}\right]f_q(\eta) = 0 \label{besseleqn} 
\ee
with the general solutions in terms of the scaling variable $x=q\eta$
\begin{eqnarray}
&&f_q(\eta) = A_q \; \eta^{\frac{5}{2}} \;\frac{J_2(x)}{x^2}
+B_q  \;\frac{x^2}{\eta^{\frac{3}{2}}} N_2(x) 
~~\mbox{for Radiation dominated} \label{RDscalingFRW} \\
&&f_q(\eta) = A_q \; \eta^{\frac{7}{2}} \;\frac{J_3(x)}{x^3}
+B_q  \;\frac{x^3}{\eta^{\frac{5}{2}}} N_3(x) ~~\mbox{for Matter dominated}
\label{MDscalingFRW} 
\end{eqnarray}
Where $ A_q~,~B_q $ are numerical constants and $ J_{p+1}(x)~;
N_{p+1}(x) $ are Bessel and Neumann functions respectively. The
constant Wronskian between $ f_q(\eta) $ and its 
complex conjugate determines that neither $ A_q $ nor $ B_q $ could
vanish. However in the asymptotic limit $ \eta \rightarrow \infty $ and 
at {\em fixed $x$}, two important simplifications occur: i) the term
proportional to $ N_{p+1}(x) $ becomes subleading and ii)  
$ q=x/\eta \rightarrow 0 $ and $ A_q \rightarrow A_0 $, thus in the
asymptotic regime for very large $ \eta $ and {\em fixed} $ x $ we find
that  the asymptotic behavior of the mode functions  
is dominated by the $J_{p+1}(x)$ contributions with a constant
coefficient $ A_0 $, i.e. asymptotically for $ \eta \gg 1 $, $ x $ fixed,
the leading contribution to the solutions are of the {\em scaling form} 
\begin{eqnarray}
&&f_q(\eta) \buildrel{\eta \gg 1 }\over= A_0 \; \eta^{\frac{5}{2}}
\;\frac{J_2(x)}{x^2} + {\cal O}\left({1 \over \eta^{\frac{3}{2}} }\right)
~~\mbox{for Radiation dominated} \label{asyRDscaling} \\
&&f_q(\eta) = A_0  \; \eta^{\frac{7}{2}} \;\frac{J_3(x)}{x^3}+ {\cal
O}\left({1 \over \eta^{\frac{5}{2}} } \right) ~~\mbox{for Matter dominated}
\label{asyMDscaling} 
\end{eqnarray}
In this asymptotic region, the absolute value of the coefficient $A_0$
is completely determined by the constraint (\ref{coeff}) resulting 
from the sum rule (\ref{sumrule}) and the integral \cite{gr} 
$$
\int_0^{\infty} {dx\over x^{2p}} \left[ J_{p+1}(x) \right]^2 =
{\sqrt{\pi} \; \Gamma(2p) \over 2^{2p+1} \;  \Gamma(2p+ \frac32 ) \;
\Gamma(p+ \frac12 )^2 } \; .
$$
We find,
\be
|A_0|^2 = 
\left\{ \begin{array}{cc}
 {15 \; \pi \over 16\; g}= 2.94524\ldots/g & ~\mbox{for Radiation dominated} \\
 {2835 \; \pi \over 512\; g}=17.39534\ldots/g &  ~\mbox{for Matter dominated} 
\end{array} 
\right.  \label{modA02} 
\ee
Figs. \ref{fig4}-\ref{fig6} display
$ g\, \eta^{-5} \,|f_q(\eta=240)|^2~;~g\, \eta^{-5} \,|f_q(\eta=400)|^2 $ and $
g|A_0~J_2(x)/x^2|^2$  
vs. $x$ respectively for 
a radiation dominated cosmology with $g=10^{-5}~;~ T_i/T_c=3$. These figures are indistinguishable from each other, furthermore
we have checked numerically that $\eta^{-\frac{5}{2}}f_{q=0}(\eta)$
approaches a constant asymptotically both for the real 
and the imaginary part in a radiation dominated FRW cosmology. Similar
results had been obtained numerically for the case of 
matter domination, in particular confirming the scaling behavior
(\ref{asyMDscaling}) with the appropriate coefficient given by 
(\ref{modA02}). In this case also   both the real and imaginary part of
$\eta^{-\frac{7}{2}}f_{q=0}(\eta)$ approach a constant 
asymptotically and  the effective mass in both
cases oscillates with small amplitude around a mean value 
given  by (\ref{masaasyntotica}) as is shown explicitly in fig. (\ref{fig7}).  

Thus a detailed numerical integration of the equations of motion
confirms the scaling {\em ansatz} and leads to the following 
conclusions:

\begin{itemize}

\item{Asymptotically the effective mass $ C^2(\eta)M^2(\eta)
\buildrel{\eta \gg 1 }\over= -{(2p+3)(2p+1) \over 4 \eta^2} $. The
vanishing of the effective mass 
leads to the sum rule (\ref{sumrule}) which suggests the scaling
ansatz (\ref{scaling}) with the power law given by (\ref{power})). The 
self-consistent solution in the asymptotic regime is given by
(\ref{RDscalingFRW}) for radiation dominated and (\ref{MDscalingFRW})
for matter dominated. Neither  
of the coefficients $A_q;B_q$ vanishes as a consequence of the
constancy of the Wronskian of $f_q$ and its complex conjugate.
However, 
the asymptotic regime $\eta \gg 1$, and $x$ fixed is completely
determined by the scaling form of the solutions given by
eqs. (\ref{asyRDscaling}), (\ref{asyMDscaling}) with the modulus
squared of the coefficient  $A_0$ determined by the sum rule and given 
by (\ref{modA02}). }

\item{In the asymptotic regime $\eta \gg 1$, $x$ fixed which is dominated by
the scaling solution we see that only the long wavelength modes 
are {\em relevant}, i.e. $q=x/\eta \ll 1$ (in units of $m_R$),
justifying keeping only the classical part $J_{cl}$ in
eq. (\ref{jotaclassical}) 
and using the cutoff $\kappa \sim m_R$ in the integral in
(\ref{jotacutoff}). }  

\item{The amplitude of the long wavelength modes $q=x/\eta$ that
dominate the scaling regime  become {\em non-perturbatively large} as 
can be seen in figs. (\ref{fig4},\ref{fig5}) and is a consequence of
the sum rule (\ref{coeff}) that fixes the amplitude of 
the coefficient of the scaling solution. The fact that the
long-wavelength modes become non-perturbatively large and dominate 
the dynamics is a consequence of the early time {\em linear
instabilities} that result in an exponential growth of these modes. } 

\item{A noteworthy feature of the scaling solution is that its phase
has {\em frozen} i.e. it became time independent and completely 
determined by the phase of $A_0$. We note that if the two linearly
independent solutions in eq.  
(\ref{RDscalingFRW},\ref{MDscalingFRW}) had the same
amplitude the phase of the mode functions will be time  and momentum
dependent. The freezing of the phase and the non-perturbative 
large amplitude of the long-wavelength modes entail that these mode
functions that originally had {\em quantum} initial conditions (as can
be seen by restoring $\hbar$ to obtain the quantum 
commutators) had become {\em classical}.  }

\end{itemize}

\subsection{Scaling corrections}
Although the leading asymptotic behavior of the mode functions is
determined by the scaling forms (\ref{asyRDscaling},
\ref{asyMDscaling}) 
the  subleading contributions determined by the terms containing the
Neumann functions $ N_{p+1}(x) $ in eqs. (\ref{RDscalingFRW},
\ref{RDscalingFRW} ) are important corrections to scaling behavior and
are required for the  self-consistency of the solutions.  

The analytic and numerical result $ C^2(\eta) M^2(\eta) \buildrel{\eta
\gg 1 }\over= -{(2p+3)(2p+1) \over 4 \eta^2} $ entails {\em two } very
stringent constraints. The first 
one leads to the cancellation between the tree level contribution and
the backreaction term $g\Sigma(\eta) =1$ to leading order in 
$C(\eta)$ and leads to the sum rule eq. (\ref{sumrule}). However the
fact that $ C^2(\eta) M^2(\eta) \buildrel{\eta
\gg 1 }\over= -{(2p+3)(2p+1) \over 4 \eta^2} $ in  
the asymptotic region when $ T_{eff}(\eta)/T_c \ll 1 $ implies that
\be
C^2(\eta) M^2(\eta)=g\int^{1}_0 {q^2\; dq} \mid f_q(\eta)\mid^2 -\;
C^2(\eta) \buildrel{\eta
\gg 1 }\over= -{(2p+3)(2p+1) \over 4 \, \eta^2}
\label{correc}  
\ee
Inserting the solutions (\ref{RDscalingFRW})-(\ref{MDscalingFRW}) in
eq. (\ref{correc}), changing integration variables to  
$ q=x/\eta $ and taking the asymptotic limit $ \eta \gg 1,  \; x $ fixed
inside the integral it is seen that the leading contribution arises
from the terms proportional to $J_{p+1}(x)$ which  
 cancel $ C^2(\eta) $.   The next order term arises
from the crossed term between the contribution proportional to the
$ J_{p+1}(x) $ and that from the $ N_{p+1}(x) $ which leads to another
sum rule
\be
\frac{2\, g \, \mbox{Re}[A_0B^*_0]}{\eta^2} \int_0^{\infty} x^2\; dx\;
J_{p+1}(x)\;N_{p+1}(x) = - \frac{(2p+3)(2p+1)}{4\, \eta^2} \label{scalecorr} 
\ee 
which  describes the asymptotic behavior of $ C^2(\eta) M^2(\eta) $. 

Using  the result\cite{gr}
$$
\int_0^{\infty} x^2\; dx\; J_{p+1}(x)\;N_{p+1}(x) =- \frac{(2p+3)(2p+1)}{16}
$$
we find that the sum-rule (\ref{scalecorr})  constrains the  value of
Re$[A_0B^*_0]$ as follows,
$$
\mbox{Re}[A_0 B^*_0] = - \frac{2}{g}
$$

\noindent thus we are led to the conclusion that the
self-consistency condition indeed requires a {\em non-vanishing
correction to scaling}. 

Eq. (\ref{scalecorr}) reveals that  $ B_0 \propto 1/\sqrt{g} $
therefore of the same order as $ A_0 $. Thus we conclude that although  
the asymptotic behavior of the mode functions is dominated by the
scaling form, the corrections to scaling embodied in the subleading 
contributions are very important for the self-consistency of the
solution, they have non-perturbative amplitudes of 
$ {\cal O}(1/\sqrt{g}) $  and as argued above the coefficients $B_q$ do
not vanish because of the Wronskian condition on the mode functions.  

\subsection{Classicality}
We argued above that the long-wavelength modes become classical in the
sense that their amplitudes become non-perturbatively 
large ${\cal O}(1/\sqrt{g})$ and their phases freeze out. An important
bonus of studying the dynamics in terms of 
a quantum density matrix is that the classicalization of
long-wavelength fluctuations can be quantified in terms of the 
probability density (functional) in field space. Just as in quantum
mechanics the probability density is the diagonal 
density matrix element in the Schr\"{o}dinger representation, i.e. 
\be
{\cal P}(\Phi)  =    
\rho[\Phi,{\Phi},t]  =  \prod_{\vec{k}} {\cal{N}}_k(t) \exp\left\{
- {\left[A_{k,R}(t)+B_k(t)\right]} \; {\vec \Phi}_k \cdot {\vec
\Phi}_{-k} \right\}  \label{proba} 
\ee
For long-wavelength modes $k<<T_i$ we can approximate $ 2
\tanh\left[\frac{\omega_k}{2T_i}\right] \approx \omega_k/T_i $ and in
terms of conformal time and the mode functions that obey the equations
(\ref{finconfeqns})-(\ref{omegafin}) we find 
\be
{\cal P}(\Phi,\eta) = \prod_{\vec{k}} {\cal{N}}_k(\eta) \exp\left\{
- \frac{C^2(\eta) \; {\vec \Phi}_k \cdot {\vec \Phi}_{-k}}{T_i \; 
|f_k(\eta)|^2} \right\}  \label{finprob} 
\ee
Thus at any given fixed (conformal) time, configurations with
 amplitude $\Phi^a_k \sim \sqrt{T_i}|f_k(\eta)|/C(\eta)$  
 (in units of $m_R$) have a
probability $\sim {\cal O}(1)$  of being represented
in the statistical ensemble. Since $ T_i \geq T_c \approx 1/g $ and in
 the asymptotic regime ($\eta \gg1~;~ x~ \mbox{fixed}$) 
$ |f_k(\eta)| \propto 1/\sqrt{g} $,   
these are {\em large amplitude} $ {\cal O}(1/g) $, {\em
long wavelength} $k \leq 1/\eta$ field  configurations. 

Thus whereas asymptotically for $ \eta \gg \eta_{NL} $ these large
amplitud long-wavelength configurations are 
represented in the ensemble with probability $ {\cal O}(1) $, in the initial
 density matrix with $ |f_k(0)| \approx 1 $  
these field configurations  are represented in the ensemble with
 probability $ \propto e^{-\frac{1}{g}}\ll 1 $ in the weak 
coupling limit. Under time evolution the Gaussian probability density
 (functional) for long-wavelength modes spreads out 
 and large amplitude long-wavelength
configurations acquire non-vanishing probabilities of being
 represented in the statistical ensemble. Thus in the asymptotic 
scaling regime, a typical field configuration found in the statistical
ensemble will have amplitude ${\cal O}(1/g)$ and the Fourier transform
of its spatial profile will be dominated by long-wavelength modes $k \ll
1/\eta$. This observation leads to a {\em semiclassical stochastic}  
description in terms of semiclassical field configurations that
describe a typical member of the ensemble for $\eta \gg \eta_{NL}$ 
\be
\Phi^a_{typ}(\vec x,\eta) \approx \sum_{\vec
k}\frac{\sqrt{T_i} \; |f_k(\eta)|}{C(\eta)}\cos[\vec k \cdot \vec x +
\delta^a_{\vec k}] 
\ee
where the phases $ \delta^a_{\vec k} $ are stochastic with a Gaussian
distribution  in order to reproduce the field correlation
function obtained 
from the Gaussian quantum density matrix given by the integrand in
(\ref{jotaclassical}). The phases $\delta_k$ represent the 
phases of the coefficients $A_q$ in the scaling solutions
(\ref{asyRDscaling})-(\ref{asyMDscaling}) which as argued above is
time 
independent. We note that a spatial translation can be absorbed into a
redefinition of the phase $\delta_k$ and the stochastic nature 
of this variable in terms of a Gaussian probability distribution
restores translational invariance in the ensemble averages of the 
semiclassical but stochastic field configurations.

\section{Consequences of scaling: Dynamical correlation length and
equation of state} 

\subsection{Dynamical correlation length}

As described in the introduction in known systems the dynamics of
phase ordering leads to the emergence of a dynamical 
correlation length $\xi(t)$ which determines the size of the
correlated regions\cite{bosecon}-\cite{bray}. This dynamical
correlation length plays  
the same role in the description of the dynamics as the static
correlation length does in {\em static} critical phenomena. 
When the correlation length becomes much larger than the typical
microscopic length scale in the system it becomes the 
only {\em relevant} length scale and much in the same way as in static
 critical phenomena a scaling regime emerges where the 
correlation length provides the natural scale for all dimensionful
quantities. The emergence of the dynamical correlation 
length in dynamical critical phenomena is revealed  by the equal time
correlation function $\langle \Phi^a(\vec x, t) \Phi^b(\vec
0,t)\rangle$. From the discussion in the previous sections, this
correlation function will be dominated by the 
long-wavelength modes, hence we use the high temperature limit $ T_i
>>k $ to find 
\be
\langle \Phi^a(\vec x, t) \Phi^b(\vec 0,t)\rangle = \delta^{a,b} \;  T_i \; 
\int \frac{k \; dk}{4\pi^2} \; \frac{\sin{kr}}{r} \;
\frac{|f_k(\eta)|^2}{C^2(\eta)} 
\label{correlationfunc}
\ee
In the scaling regime when the mode functions are of the form
(\ref{scaling}) and multiplying by the coupling 
constant $\lambda$ to write the result in a more familiar manner we find
\bea
&&\lambda \langle \Phi^a(\vec x, t) \Phi^b(\vec 0,t)\rangle =
\delta^{a,b} \; g \; D(z) \nonumber \\ 
&& D(z) = \frac{1}{2z}  \int_0^{\infty} {x \; dx}~ {\sin{2xz}}~|{\cal
F}(x)|^2 ~~; ~~ z=\frac{r}{2\eta}  \label{correfuncofz} 
\eea
where we have introduced the {\em scaling ratio} $z=r/2\eta$. Figures
\ref{fig8}, \ref{fig9}  show $gzD(z)$ as
a function of $z$ for $\eta=240,400$ (in units of $m_R$) respectively
for radiation dominated cosmology obtained from 
the integration of the mode functions $f_k(\eta)$, and figure 11 shows
$ gzD(z) $ for the matter dominated case for $ \eta=400 $. A remarkable
feature of the correlation function $ D(z) $ is that it 
{\em becomes of the order $g$ } for $r>2\eta$ (i.e, $ z>1 $) in both
cases, obviously as a 
result of causality. This result leads to 
the conclusion that the dynamics of phase ordering is described in the
scaling regime as the growth of correlated 
regions of comoving size $2\eta$, the reason for the factor
$2$ is that one edge of this region is localized at the point
$\vec 0$ and the other at $\vec r$ and the boundary 
of this correlated region recedes at the speed of light.  

 Hence we obtain that the {\em comoving} dynamical length scale that
 determines the typical (comoving) size of a correlated domain is
 given by 
\be
\xi_{com}(\eta) = \eta \label{comoxi}
\ee 
The {\em physical} dynamical correlation length is given by

\be
\xi_{phys}(\eta) = C(\eta) \; \eta = d_H(\eta) \label{physixi}
\ee

\noindent with $d_H(\eta)$ the size of the causal horizon. Hence we
conclude that the dynamical correlation length 
is exactly the size of the causal horizon and within one horizon there
is exactly one correlated region within which 
$\langle \vec \Phi \cdot \vec \Phi \rangle \approx m^2_R/\lambda_R$ i.e,
the mean square root fluctuation of the field is probing the  broken
symmetry ground state. This result has been obtained via the full
numerical evolution of the mode functions but can be understood
analytically as a consequence of the scaling solutions (\ref{asyRDscaling},
\ref{asyMDscaling}) that dominate the asymptotic regime for 
long-wavelengths. Replacing ${\cal F}(x)$ in (\ref{correfuncofz}) by
the scaling form of the solutions given by
(\ref{asyRDscaling})-(\ref{asyMDscaling}) in the radiation and matter
dominated cases we find  
\be
g\,z\, D(z) = \left\{ \begin{array}{cc}
 &(1-z^2)^3\, F[\frac{1}{2} ,\frac{5}{2};4;1-z^2]\Theta(1-z)
~\mbox{for Radiation dominated} \\ & \\
 & (1-z^2)^5\, F[\frac{1}{2} ,\frac{7}{2};6;1-z^2]\Theta(1-z)
~~~\mbox{for Matter dominated}  
\end{array} 
\right.  \label{scalingcorrfunc}
\ee 
where $ F[a,b;c;z] $ is the hypergeometric function. A numerical
evaluation of these expressions agrees to a very high 
level of precision with the result obtained above from the numerical
integration.   

Moreover, these particular hypergeometric functions can be expressed
in terms of complete elliptic integrals \cite{prud}
\bea
(1-z^2)^3 \, F[\frac{1}{2} ,\frac{5}{2};4;1-z^2] &=& {32 \over 15 \, \pi}
\left[ z^2 (z^2 -9 ) K(\sqrt{1-z^2}) +(3 + 7 z^2 -2
z^4)E(\sqrt{1-z^2}) \right] \quad , \cr \cr
(1-z^2)^5\, F[\frac{1}{2} ,\frac{7}{2};6;1-z^2]&=& {512 \over 945 \,
\pi }\,(1-z^2)^2 \left[ (45 z^8 -7z^6 +22 z^4 -11z^2 +15)
K(\sqrt{1-z^2}) \right. \cr \cr
&-&\left. (15 z^8 +31 z^6 +12 z^4 -17 z^2 +23)  E(\sqrt{1-z^2}) \right] \quad ,
\eea
where $ K(k) $ and $ E(k) $ are complete elliptic integrals of the
first and second kind, respectively.

The power spectrum in the asymptotic scaling regime is dominated by
the long-wavelength modes and is given by 
\be 
S(k,t) = \frac{1}{N} \langle \vec{\Phi}_{\vec k}(t)\cdot {\vec
\Phi}_{-\vec k}(t) \rangle = {\cal Z} \, \frac{\eta^3}{g} 
\; \left[ \frac{J_{p+1}(k\eta)}{(k\eta)^{p+1}} \right]^2
\ee
\noindent where $ p=1,2 $ for radiation dominated or matter dominated,
respectively and where $ {\cal 
Z} $ is a constant of order one. At long times $\eta \gg 1$ this power
spectrum becomes  strongly peaked at $k=0$ and receives contribution only
from a narrow region $ k \leq 1/\eta $ as
evidenced by figs. (\ref{fig4}-\ref{fig6}). Since the sum rule
constrains the total integral of the power spectrum to be 
\be
\int k^2 \; dk \; S(k,t) =\mbox{constant} \label{constra}
\ee
\noindent we are led to the conclusion that at asymptotically long
times the power spectrum is sharply peaked at 
$k=0$ becoming asymptotically a delta function. This is the signal of
the formation of a zero momentum condensate, much in the same manner
as observed in the case of Minkowski space-time\cite{bosecon,destri}
and also in dynamical critical phenomena in condensed matter
systems\cite{nosreviews,bray}.   

\subsection{Equation of state}

The expectation value of the energy momentum tensor in the
non-equilibrium density matrix is of the fluid form in 
terms of the energy density $\varepsilon$ and pressure $P$ given by
\bea
&&\frac{\lambda}{N}\varepsilon = \frac{1}{2}\left[m^4_0 -
\left(\frac{\lambda}{2N}\langle \vec\Phi \cdot \vec\Phi \rangle
\right)^2 \right]\ 
 + \frac{\lambda}{2} \int d^3k \; \coth\left[\frac{\omega_k}{2T_i}\right]\left[
|\dot{\varphi}_k(t)|^2+ W^2_k(t)|\varphi_k(t)|^2 \right] \label{enerdens}\\
&& \frac{\lambda}{N}(P+\varepsilon) = \lambda \int d^3k \; 
\coth\left[\frac{\omega_k}{2T_i}\right]\left[ 
|\dot{\varphi}_k(t)|^2+ \frac{k^2}{3a^2(t)}|\varphi_k(t)|^2 \right]
\label{pressdens} 
\eea
where $\varphi_k(t)$ obeys the equations of motion (\ref{modesk}) with
$W^2_k(t)$ given by eq. (\ref{timefreqs}) and we had 
set to zero the coupling to the Ricci scalar. Using
the equations of motion (\ref{modesk}) and the definition of the
self-consistent 
frequencies $W^2_k(t)$ (\ref{timefreqs}), it can be easily verified that 
$\varepsilon$ and $P$ satisfy the covariant conservation law
\be
\dot{\varepsilon}(t)+3 \, \frac{\dot{a}}{a}\, [\varepsilon(t) + P(t)]= 0
\label{conservation}  
\ee
Passing on to conformally rescaled mode functions and conformal time
and writing the asymptotic forms of the scale 
factor in radiation dominated and matter dominated dominated cosmologies as
\be
C(\eta) = \left( { \eta \over 2 } \right)^p \quad \mbox{with}~ p=1
~\mbox{for Radiation dominated} ~ ; ~ p=2 ~\mbox{for Matter dominated}
\label{scalefacasy} 
\ee

\noindent we will focus our discussion on the behavior of the pressure
and the energy density in the scaling regime  
wherein the conformally rescaled mode functions in conformal time are
given by eq.  
(\ref{asyRDscaling}, \ref{asyMDscaling}), which can be handily written as
\be
f_k(t) = A_0  \; \eta^{p+3/2}  \;  \frac{J_{p+1}(x)}{x^{p+1}}
\ee 
in either case. The sum rule (\ref{masaasyntotica}) with vanishing
coupling to the Ricci scalar leads to the 
following identity in the asymptotic regime
\be
C^2(\eta)\left[-m^2_0 +\frac{\lambda}{2N}\langle \vec\Phi \cdot
\vec\Phi \rangle-\frac{C''(\eta)}{C^3(\eta)}\right] \rightarrow
-\frac{(2p+3)(2p+1)}{4\eta^2} + {\cal O}(1/\eta^4) \label{suma} 
\ee
up to perturbatively small corrections. The leading term in this sum
rule is given by the sum rule (\ref{sumrule}) 
which upon using the asymptotic scaling mode functions
(\ref{asyRDscaling})-(\ref{asyMDscaling}) can be written in the 
following  compact form
\be
|A_0|^2 \; g  \int dx \left[\frac{J_{p+1}(x)}{x^p} \right]^2
= 2^{-2p} \label{compactform} 
\ee
Using this result and after some straightforward algebra we find that
the asymptotic behavior of the energy density and pressure are given by
\bea
\frac{\lambda}{N}\varepsilon  =
\frac{|A_0|^2\; g\; 2^{2p} }{C^2(\eta) \; \eta^2}  \int
x^2dx && \left\{ \left[(\frac{1}{2}-p)\frac{J_{p+1}(x)}{x^{p+1}}+
\frac{J'_{p+1}(x)}{x^p}\right]^2 + 
x^2\left[\frac{J_{p+1}(x)}{x^{p+1}} \right]^2
\right\}\label{enerdensofx} 
\eea
\bea
\frac{\lambda}{N}(\varepsilon+P) = \frac{|A_0|^2\; g\;
2^{2p+1}}{C^2(\eta) \;\eta^2} \int x^2dx && \left\{ 
\left[(\frac{1}{2}-p)\frac{J_{p+1}(x)}{x^{p+1}}+
\frac{J'_{p+1}(x)}{x^p}\right]^2 + 
\frac{x^2}{3}\left[\frac{J_{p+1}(x)}{x^{p+1}} \right]^2
\right\} \label{pressureofx} 
\eea
\noindent where the prime stands for derivative with respect to the
argument. The integral is independent of conformal 
time and the only dependence on $ \eta $ is in the prefactors, the product 
\be
C^2(\eta) \; \eta^2 \propto \left\{ \begin{array}{cc}
 &C^4(\eta)~\mbox {for Radiation dominated} \\
& \\
 & C^3(\eta) ~\mbox{for Matter dominated} 
\end{array} 
\right. 
\ee
thus we find that in either case 
\be
\varepsilon \propto \rho_{back}(\eta) \label{identity}
\ee
\noindent where $\rho_{back}(\eta)$ is the {\em background} energy
density in either radiation dominated ($\rho_{back}(\eta) \propto 
C^{-4}(\eta)$) or matter dominated ($\rho_{back}(\eta) \propto
C^{-3}(\eta)$). Since $\varepsilon$ and $P$ satisfy the covariant 
conservation equation and so do the background energy density and
pressure we conclude that the fluid resulting 
from the fluctuations of the scalar field in the asymptotic scaling
regime {\em obeys the same equation of state as 
the background fluid}. Indeed careful  evaluation of the integrals
using the properties of the Gamma function \cite{gr} as analytic functions of
the variable $p$ lead to the following remarkable results 
\be
\frac{\lambda}{N} \; \varepsilon(\eta) \;C^2(\eta) \;\eta^2 = 3 \;p \;
\frac{p+1/4}{p-1} \label{resultenergy} 
\ee
\be 
\frac{\lambda}{N}[\varepsilon(\eta)+P(\eta)] \; C^2(\eta) \;\eta^2 =
2 \;(p+1) \; \frac{p+1/4}{p-1} \label{resultpress} 
\ee
\noindent leading to one of the important results of this article
\be
\frac{P}{\varepsilon}= \frac{1}{3} \left(\frac{2}{p}-1\right) =
\left\{ \begin{array}{cc} 
 &\frac{1}{3}~\mbox {for Radiation dominated} \;(p=1) \\ & \\
& 0 ~\mbox{for Matter dominated} \; (p=2) \end{array} 
\right. \label{povere}
\ee

The simple poles at $p=1$ in
eqs. (\ref{resultenergy})-(\ref{resultpress}) reflect ultraviolet 
logarithmic singularities in the integrals which have been evaluated 
with an upper limit taken to infinity. However we remark that the
scaling form requires that the upper limit be 
of order $x \ll 1/\eta$ so that the coefficient of the Bessel
functions can be replaced by momentum independent constants 
leading to true scaling solutions. A numerical evaluation of the
integrand in (\ref{enerdensofx})-(\ref{pressureofx}) 
reveals that these are strongly peaked near $x=0$ and most of the
contribution arises from the interval $ x\leq 5 $ even 
for $p=1$. The results (\ref{resultenergy})-(\ref{resultpress}) are the
analytic continuation of 
these integrals in the variable $p$ taking the upper  cutoff to
infinity. The variable $p$ thus acts as a regulator much 
in the same way as analytic or dimensional regularization and the equation of
state is insensitive to the regularization procedure.  

Defining $ \varepsilon = \delta \rho $ with $ \varepsilon $ the energy
density given by eq.(\ref{enerdensofx}) and 
$ \delta \rho $ as the contribution to the energy density from the
fluctuations of the scalar field, the fact 
that $ \delta \rho /\rho_{back}= $constant i.e, independent of time is
the statement that the power spectrum for  
the density fluctuations is of the Harrison-Zeldovich form, i.e.,
scale invariant, for long-wavelength scalar density
perturbations\cite{reviews,kibble}.

\subsection{Universality of scaling}
Although we have studied radiation and matter dominated cosmologies
separately while the most physical scenario involves 
a smooth transition between the two regimes we now argue that if the
transition between the two regimes occurs over a 
very long time scale our analysis leads to the conclusion that scaling
will be a robust feature in the regimes in which 
the scale factor is dominated by a power law. In the asymptotic
regime, when the effective time dependent mass vanishes, 
the sum rule
\be
g\int^{1}_0 {q^2  \; dq} \mid f_q(\eta)\mid^2 = C^2(\eta) \label{suma2}
\ee
is fulfilled for an arbitrary scale factor, not necessarily a power
law, however if $C(\eta)$ is {\em not} a power 
law, the solution is {\em not} of the scaling form. 

Consider the scale factor in the general case in which the background fluid
has a radiation and a matter component as given by (\ref{Cofeta}),i.e,

\be
C(\eta) = 1+ H_i \; \eta + \frac{H^2_i~ r_i}{4(1+r_i)} \; \eta^2
\ee
 
\noindent in a realistic scenario the ratio of the initial matter
density to the initial radiation density is $ r_i \ll 1 $. Therefore 
in the regime $\frac{1}{r_i} \gg H_i \; \eta \gg 1 $ the scale factor is
approximately that of a radiation dominated cosmology 
and  therefore a power law, hence the sum rule above (\ref{suma2})
leads to the scaling form of the solution. During the time scales of
the crossover between radiation and matter domination $r_iH_i \eta
\approx 1$ the scale factor is {\em not} a pure power law. The sum 
rule above {\em still holds} but it does not entail a scaling form for
the mode functions. However for $H_i\eta r_i \gg 1$ 
again the scale factor is a power law now describing a matter
dominated cosmology and the sum-rule again leads to 
a scaling form for the solution in the long-wavelength limit. However
when the new scaling form emerges the solution  
{\em does not} depend on the past history, i.e. on the violations of
scaling during the crossover or the previous scaling 
solution in the radiation dominated regime. The reason for this is
that the sum rule above is {\em local} in time and 
fixes the absolute value of the coefficient $A_0$ and the scaling form
in terms of the Bessel function, however the 
{\em phase} of the coefficients $A_q$ and certainly of $A_0$ will
depend on the history and will {\em not} be determined 
by the sum-rule. However in the semiclassical stochastic description
advocated in the previous section this phase is 
treated as a stochastic variable with a Gaussian distribution
function. Hence all of the initial information and the history
contained in the phases of the scaling solution can be treated
stochastically for the long-wavelength components. 

\section{Comparison with the $O(N)$ non linear sigma model:} There are
many similarities but also many differences with  
the results obtained in the case of the {\em classical} non-linear
sigma model in refs.\cite{turok,durrer}. The 
non-linear sigma model describes the non-linear interactions of
Goldstone bosons in a broken symmetry phase. To begin 
with let us establish an important difference: the non-linear sigma
model {\em cannot} be used to describe the {\em 
phase transition} because the fields are constrained to the vacuum
manifold. Hence all of the details of 
the dynamics of the phase transition, the linear instabilities, growth
of long-wavelength modes, the different time 
scales and regimes and the explanation of why the power spectrum is
peaked at long-wavelength as a consequence of 
the early exponential growth of long-wavelength fluctuations simply
cannot be captured by the non-linear sigma model.  
These details that are deeply dependent on the details of the phase
transition have important consequences: the amplitude of the mode
functions in the scaling regime is {\em non-perturbatively large} in
the weak coupling limit since $A_0 \propto 1/\sqrt{g}$. This amplitude
is not arbitrary, if 
$ \hbar $ is restored,  the initial  amplitudes of the mode functions
are of  $ {\cal O}(\sqrt{\hbar}) $, the non-perturbative 
growth of the amplitude of long-wavelength fluctuations is a
consequence of the linear exponential instabilities which 
are a dynamical hallmark of the phase transition. Furthermore the
corrections to scaling embodied in the contribution 
from the Neumann functions in
eqs.(\ref{RDscalingFRW})-(\ref{MDscalingFRW}) have very precise
coefficients $B_q$ since 
these are constrained by the Wronskian condition {\em and the
self-consistency condition} (\ref{scalecorr}). Both the 
coefficients $A_q~;~B_q$ carry information from the initial conditions
because of the constancy of the Wronskian, the 
long wavelength limit $A_0~;~B_0$ are of order ${\cal O}(1/\sqrt{g})$
as a consequence of the self-consistency 
conditions (\ref{modA02})-(\ref{scalecorr}) and the phases can only be
determined from the numerical evolution.  

There are also very important similarities: in the asymptotic regime
the scaling form of the solution
(\ref{asyRDscaling})-(\ref{asyMDscaling}) is the {\em same} in the
model studied here and in the $O(N)$ non-linear sigma
model\cite{turok,durrer}. In the non-linear sigma model the 
coefficient of the Bessel function is fixed  by  the
length of the $O(N)$ vector, 
 this length is interpreted as the vacuum expectation value of the
field. In fact in the model studied here the 
sum-rule (\ref{sumrule}) can be identified with the same
constraint. The similarity of the asymptotic scaling solutions 
is a consequence of the fact that eq.(\ref{sumrule}) is local as
discussed above. 

The description of the non-equilibrium dynamics via the evolution of
the quantum density matrix leads to a clear interpretation of the
emergence of a semiclassical stochastic description of long-wavelength
fluctuations at long times. Our analysis thus provides 
a consistent {\em microscopic, quantum field theoretical derivation} for the
classical stochastic treatment of the Goldstone modes in the
asymptotic region which was used in
references\cite{turok,durrer}. Furthermore, our detailed analysis 
quantifies the time scales for which such semiclassical stochastic
description is valid and goes beyond providing corrections to scaling.  

In ref.\cite{turok} the energy density of the scalar field in the 
asymptotic scaling regime was found to be proportional to that of the
background in the  matter dominated case and differing 
from that of the background by a weak logarithmic dependence on $\eta$
in the radiation dominated regime. The logarithmic 
dependence has a counterpart in our result in the form of the pole in
the energy density and pressure as $p\rightarrow 1$ 
in the expressions (\ref{resultenergy})-(\ref{resultpress}) obtained via
an analytic continuation in the variable $p$, as we 
pointed out the equation of state is independent of the
regularization. Furthermore we emphasized that both the energy 
density and the pressure are dominated by modes $k<<1/\eta$ i.e. {\em
superhorizon} modes.

\section{Conclusions and further questions}

In this article we have studied in detail the non-equilibrium dynamics
 of a symmetry breaking phase transition in a 
 spatially flat radiation and or matter dominated FRW
 background. Anticipating the necessity for a non-perturbative  
treatment, we studied the linear sigma model with a scalar field in
 the vector representation of $O(N)$ in leading order in 
the large $ N $ limit. The Liouville equation for the quantum density
 matrix is solved in this limit from which we extract 
the necessary correlation functions to leading order. The advantage of
 working with a density matrix is that this description allows a clear
 interpretation of the emergence of a semiclassical description in
 terms of a field probability distribution functional.

The main goal of our study is to provide a thorough
understanding the process of phase ordering beginning from a state of
local thermodynamic 
equilibrium at an initial temperature larger than the critical, the
cosmological expansion triggers the phase transition when the
effective time dependent temperature falls below the critical. The key
issues that we address in this article 
are the emergence of a scaling regime and of a  characteristic
dynamical correlation length that determines the spatial extent of the
correlated regions of broken symmetry and the consequences of such
scaling behavior upon the equation of state of the field fluctuations.  

The non-equilibrium dynamics after the phase transition is
characterized by three distinct time scales: during the early  
stage after the phase transition the dynamics is dominated by the
exponential growth of long-wavelength fluctuations 
associated with  spinodal instabilities. This is essentially a linear
regime in which the backreaction can be ignored  
for weak self-coupling and an analytic description is available. An
intermediate time scale is defined when the backreaction from the
self-consistent mean-field 
is comparable to the tree level terms in the equations of motion and
determines the onset of non-perturbative and non-linear 
dynamics which must be studied numerically. A third, asymptotic time
scale reveals the emergence of a scaling regime for radiation or
matter dominated FRW backgrounds. This stage is dominated by 
 large amplitude  fluctuations with wavelengths of order of or larger
than the causal horizon. This regime is characterized by 
a dynamical  physical correlation length $\xi_{phys}=d_H(t)$ with
$d_H(t)$ the size of the causal horizon and the onset of a
non-equilibrium condensate at zero momentum. The dynamical
correlation 
length determines the size of the correlated domains inside which the
field fluctuations probe the broken symmetry states, hence there 
is exactly one correlated domain per causal horizon. The field
correlations vanish for distances larger than $d_H(t)$ by causality. 
In this regime the phases of the long-wavelength quantum mode
functions become time independent and their amplitude becomes
non-perturbatively large, the approach via the density matrix reveals
the emergence of a semiclassical but stochastic description and
provides a microscopic justification for a semiclassical stochastic
treatment in the asymptotic regime.    
 
A remarkable corollary of the scaling solution of the equations of
motion is that in the asymptotic regime the equation of state 
of the fluid described by the fluctuations of the scalar field is the same as
that of the background fluid that drives the dynamics of the 
scale factor. An important consequence of this behavior of the fluid
is that $\delta \rho /\rho_{back}=$constant which results in 
a Harrison-Zeldovich spectrum of scalar density perturbations for
superhorizon wavelengths\cite{reviews,kibble}.  

The self-consistency of the equations of motion entails very precise
corrections to scaling. We argue that the 
scaling solution is a universal feature of a scale factor that is a
power law and is independent of the crossover between the 
radiation and matter dominated regimes. 

Some important aspects remain to be explored further. A thorough
investigation of cosmological perturbations has been performed  
in ref.\cite{durrer}. An important aspect of the models under
consideration is that the perturbations in the energy momentum 
tensor are {\em non-linear} in terms of the field fluctuations, unlike
the case of scalar field perturbations in the inflationary 
stage. The correction to the energy density is given by the
expectation value of the energy momentum tensor of the fluctuations
and even to leading order in the large $ N $ limit, this is a quadratic form
in terms of the fluctuations [see eq.(\ref{enerdens})]. This 
feature has an important consequence in the spectrum of primordial
density perturbations: decoherence effects tend to suppress the 
acoustic peaks at large angular momentum $ l $ (small angular
resolution) (see \cite{durrer} for a clear treatment in the non-linear
sigma model). As we have argued above, there are very precise and
important corrections to scaling that are manifest for subhorizon
wavelengths, i.e. $ k \gg 1/\eta $. An important 
possibility is that these corrections to scaling affect the
correlations of the energy momentum tensor and therefore the power
spectrum of scalar density  perturbations on small angular
scales. This would arise 
from the interference effects between the scaling contributions and
those of the scaling corrections, much in the same manner as the 
next-to-leading contribution in the sum rules discussed above. As we
have highlighted these interference effects are necessary for  
the self-consistency of the method leading to a precise form of the
scaling corrections. These corrections had not been taken into account 
in the analysis of the power spectrum at small angular scales and
whether they lead to significant corrections and or a dramatic change 
of the picture of decoherence effects in causal perturbations is an
open question that merits careful study.   

Another important aspect that requires further  understanding is the
description of the non-equilibrium dynamics in 
the case of a single scalar field which cannot be addressed reliably
in the large $ N $ expansion. In the large $ N $ limit an important 
feature of the phase diagram of the theory is that the coexistence and
the spinodal line merge as a consequence of the Ward identity
associated with the continuous symmetry\cite{inflation,nuestros} and
the long-wavelength physics  described by the dynamics of 
Goldstone bosons. Static renormalization group arguments in scalar
field theories lead to a Maxwell constructed free energy with  
a region of phase coexistence that joins the minima of the free energy
and no metastable region\cite{polonyi,wetterich} for all values of $ N
$ with no basic difference between the cases of a single scalar field
or a multiplet of fields.  
 It is important to find a consistent non-perturbative approximation
scheme for $ N=1 $  to study the {\em dynamics} in this case. 
Another important aspect to be studied further is whether the scaling
solution survives $1/N$ corrections, in particular it is clear 
that in next to leading order in the large $ N $ approximation another
time scale associated with collisional processes should emerge and 
the relevant question is whether this microscopic scale will modify
substantially the scaling solution and in particular the equation of 
state. These and other related questions are currently under consideration. 

\section{Acknowledgements:} 
 D. B. thanks  the N.S.F for
partial support through grant awards: PHY-9605186 and INT-9815064 and
LPTHE, University of Paris VI and VII, for warm 
hospitality, H. J. de Vega thanks the Dept. of Physics at the Univ. of
Pittsburgh for hospitality. The authors thank the Institute 
for Nuclear Theory at the University of Washington for hospitality
during this work.  We thank the CNRS-NSF exchange programme  for
partial support.  


\begin{figure}
{\epsfig{file=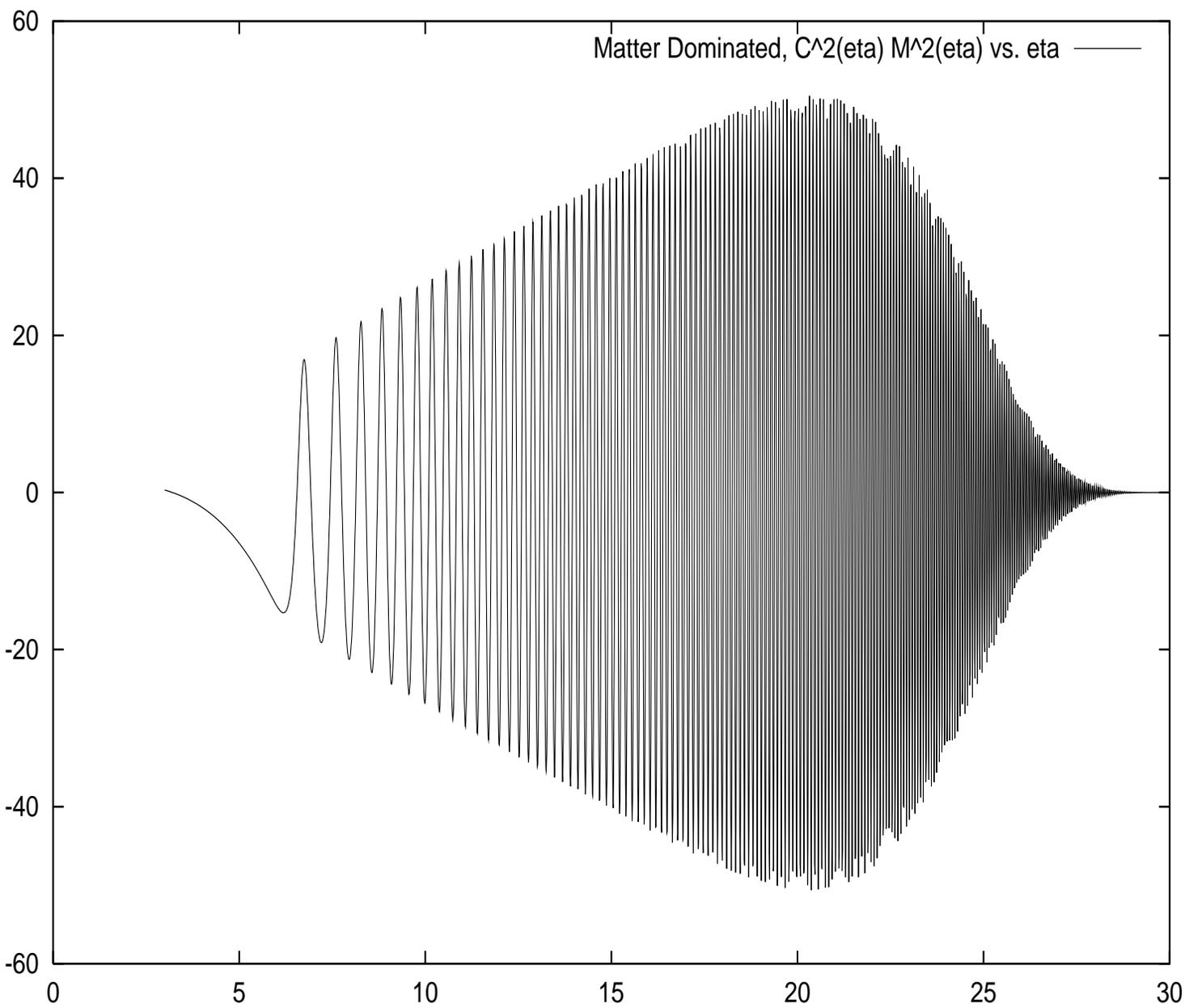,width=17cm,height=20cm}}
\caption{ $ C^2(\eta) M^2(\eta) $ vs. $ \eta $ (conformal time in units of
$ m^{-1}_R $) for $ \frac{T_i}{T_c}=1.1 \, , g=10^{-4}$. Matter
dominated universe. \label{fig3md} } 
\end{figure}

\begin{figure}
{\epsfig{file=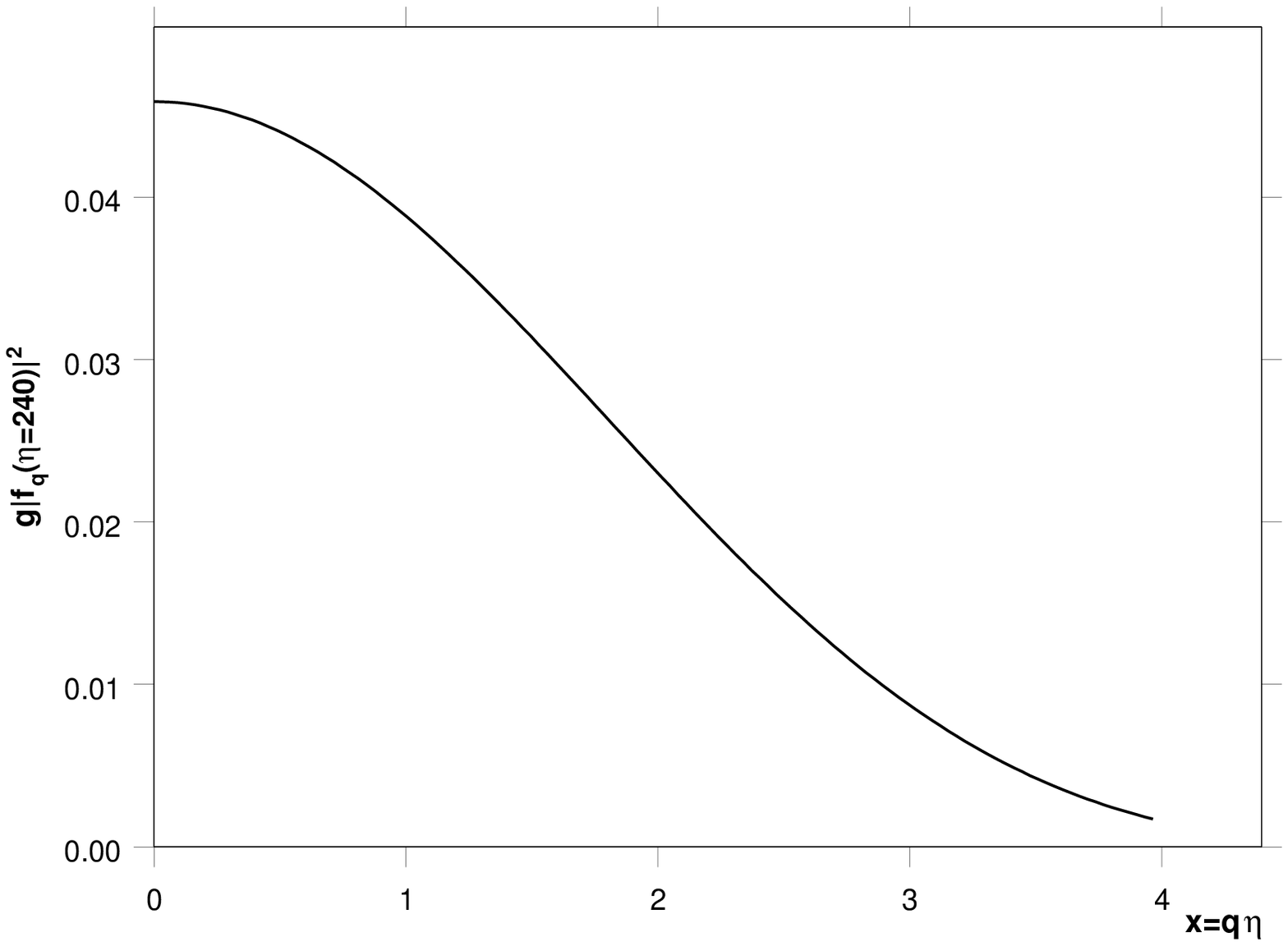,width=7in,height=8in}}
\caption{ $ g \, \eta^{-5} \, |f_q(\eta=240)|^2 $ vs. $x=q\eta$  for
$\frac{T_i}{T_c}=3$, $g=10^{-5}$. Radiation dominated universe.
\label{fig4} } 
\end{figure}

\begin{figure}
{\epsfig{file=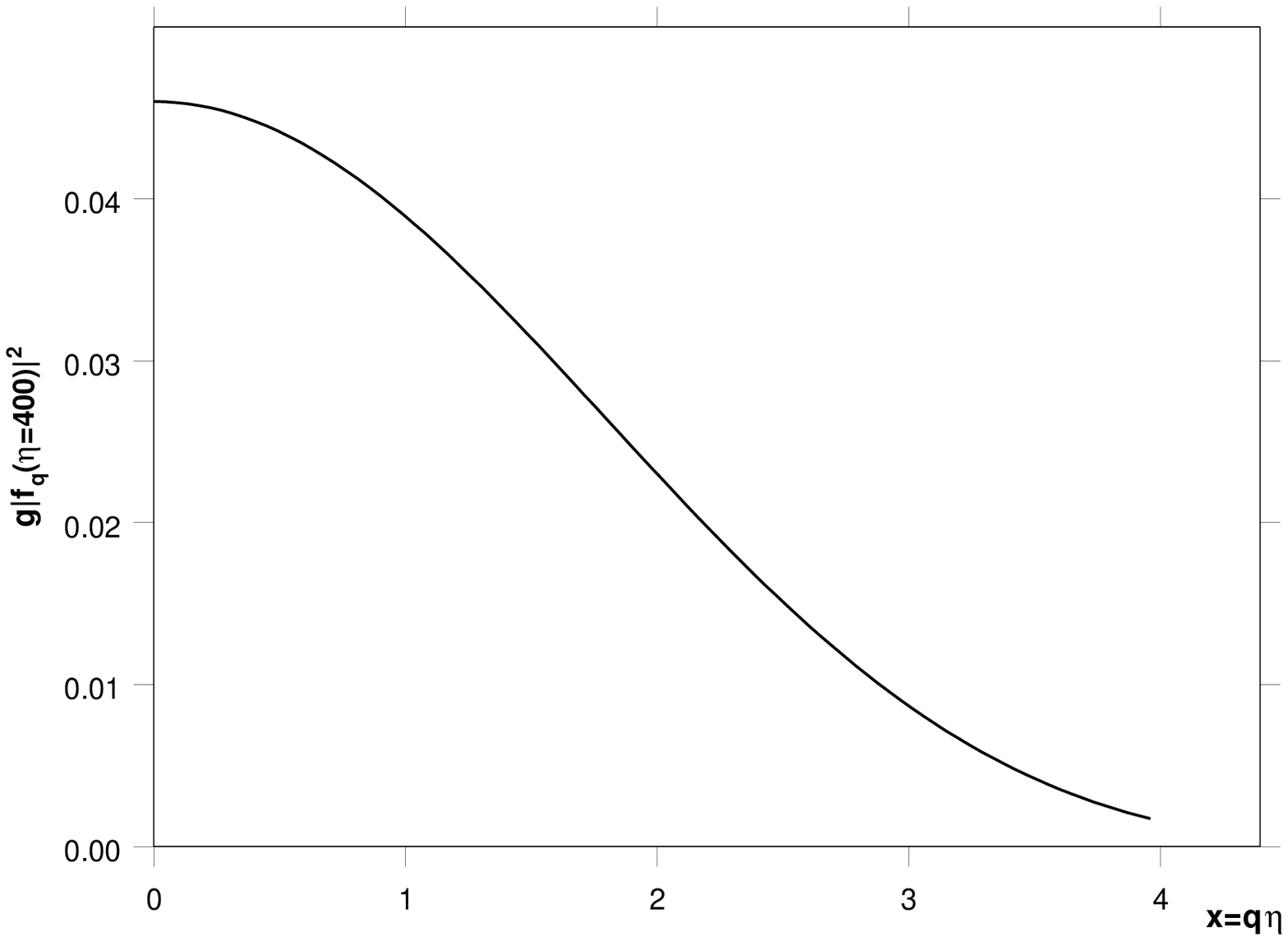,width=7in,height=8in}}
\caption{ $g\, \eta^{-5} \,|f_q(\eta=400)|^2$ vs. $x=q\eta$  for
$\frac{T_i}{T_c}=3$, $g=10^{-5}$.  Radiation dominated universe.
\label{fig5} } 
\end{figure}

\begin{figure}
{\epsfig{file=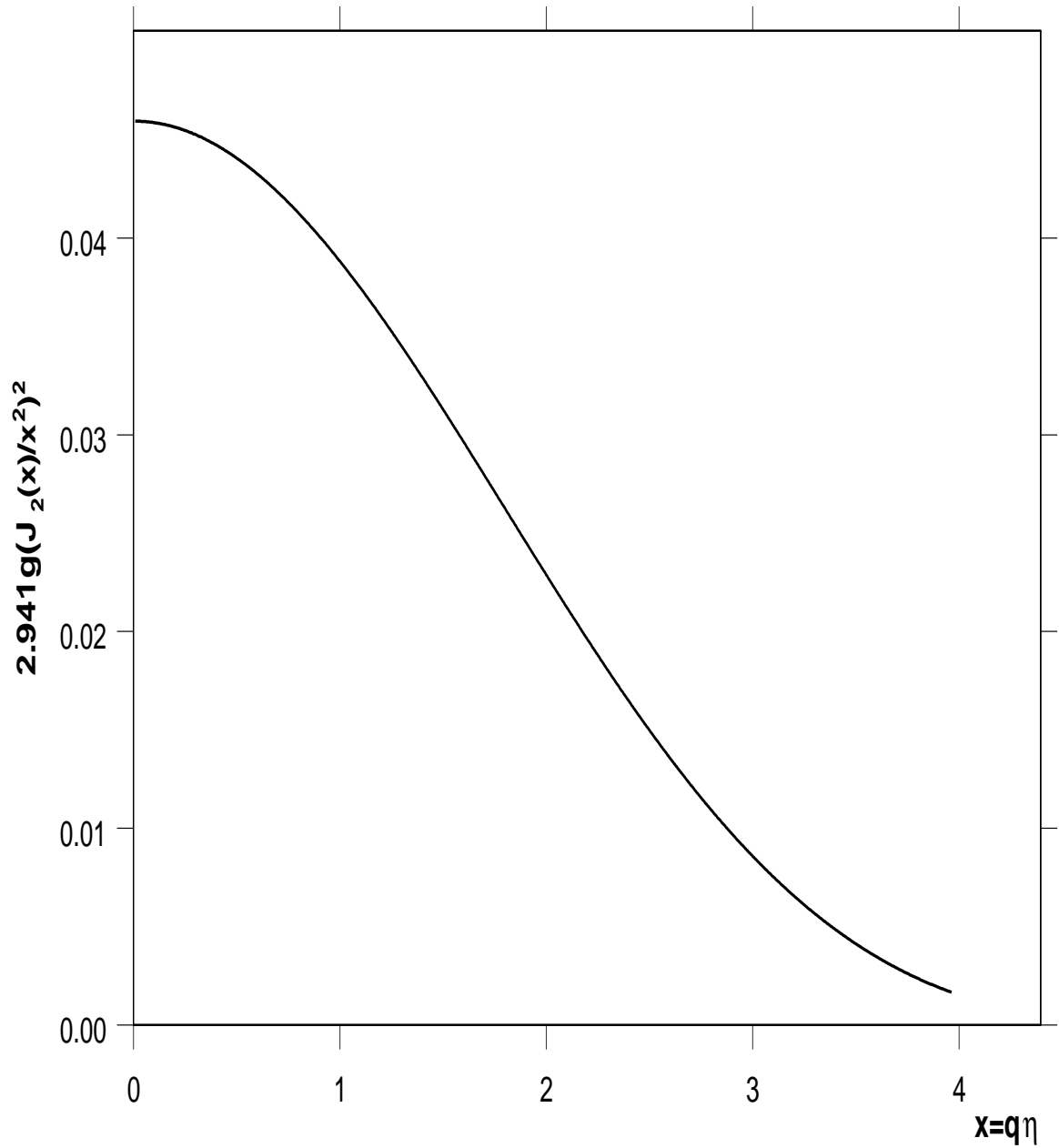,width=7in,height=8in}}
\caption{ $ g|A_0 J_2(x)/x^2|^2 $ vs. $ x=q\eta $  for $
\frac{T_i}{T_c}=3 $, $ g=10^{-5} $. Radiation dominated universe.
\label{fig6} } 
\end{figure}

\begin{figure} 
{\epsfig{file=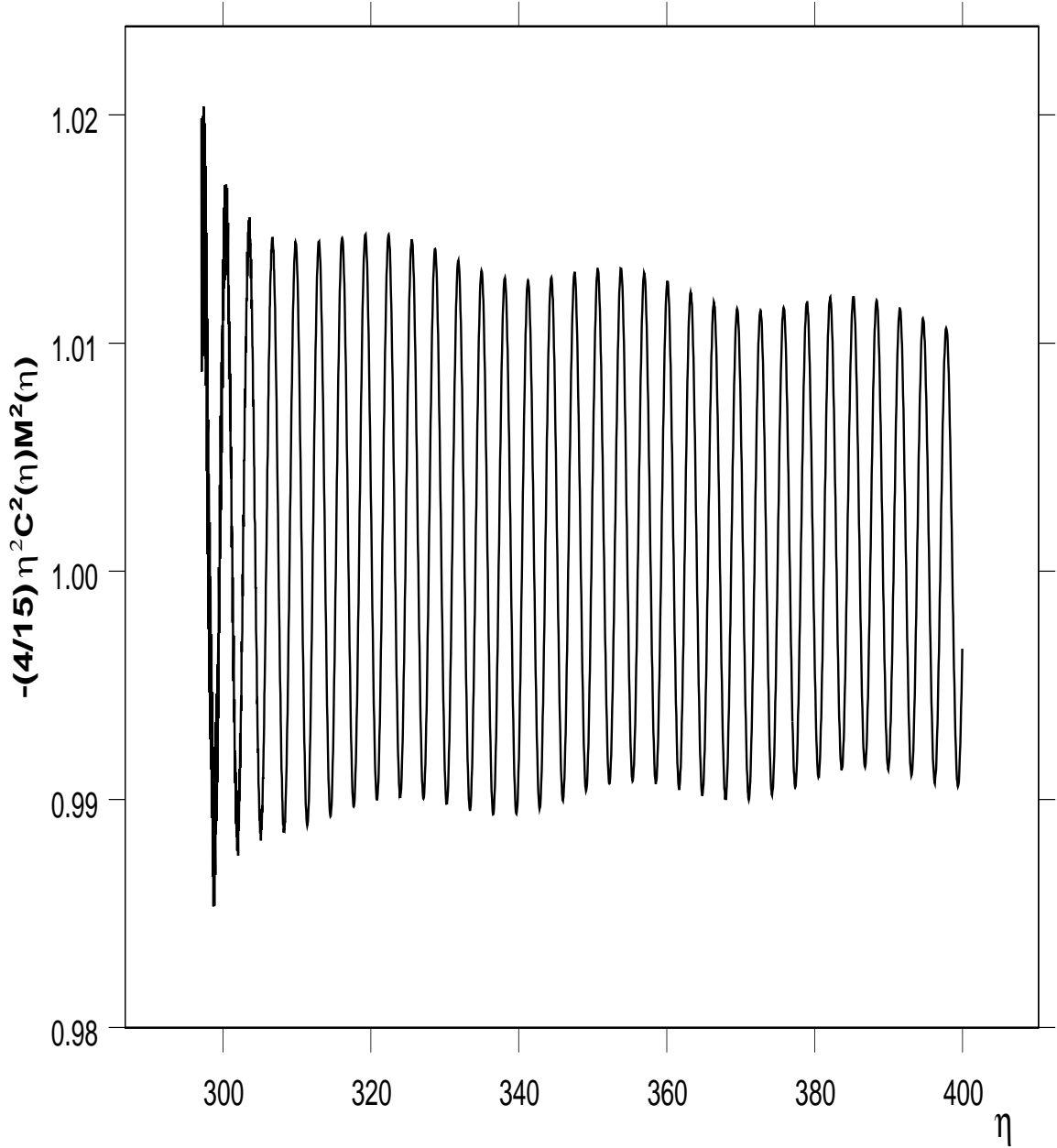,width=7in,height=8in}}
\caption{ $(-4/15)\eta^2 C^2(\eta) M^2(\eta)$ vs. $\eta$ (in units of
$m^{-1}_R$  for $\frac{T_i}{T_c}=3$, $g=10^{-5}$. Radiation dominated
universe. \label{fig7} } 
\end{figure}

\begin{figure}
{\epsfig{file=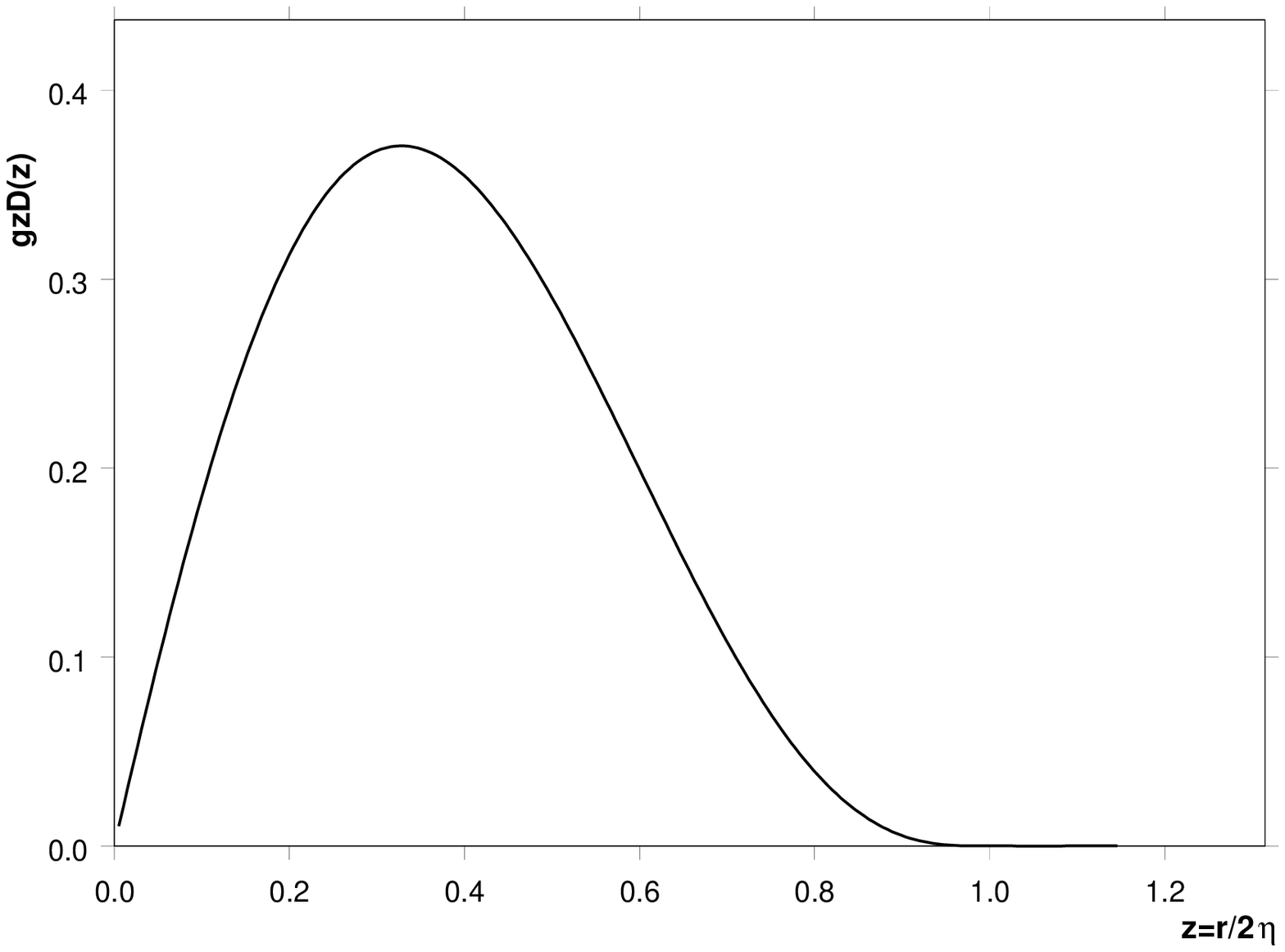,width=7in,height=8in}}
\caption{ $ gzD(z) $ vs. $ z=r/2\eta $ at $ \eta=240 $ (in units of $
m_R $)  for $ \frac{T_i}{T_c}=3 $, $ g=10^{-5} $. Radiation dominated universe.
\label{fig8} } 
\end{figure}

\begin{figure}
{\epsfig{file=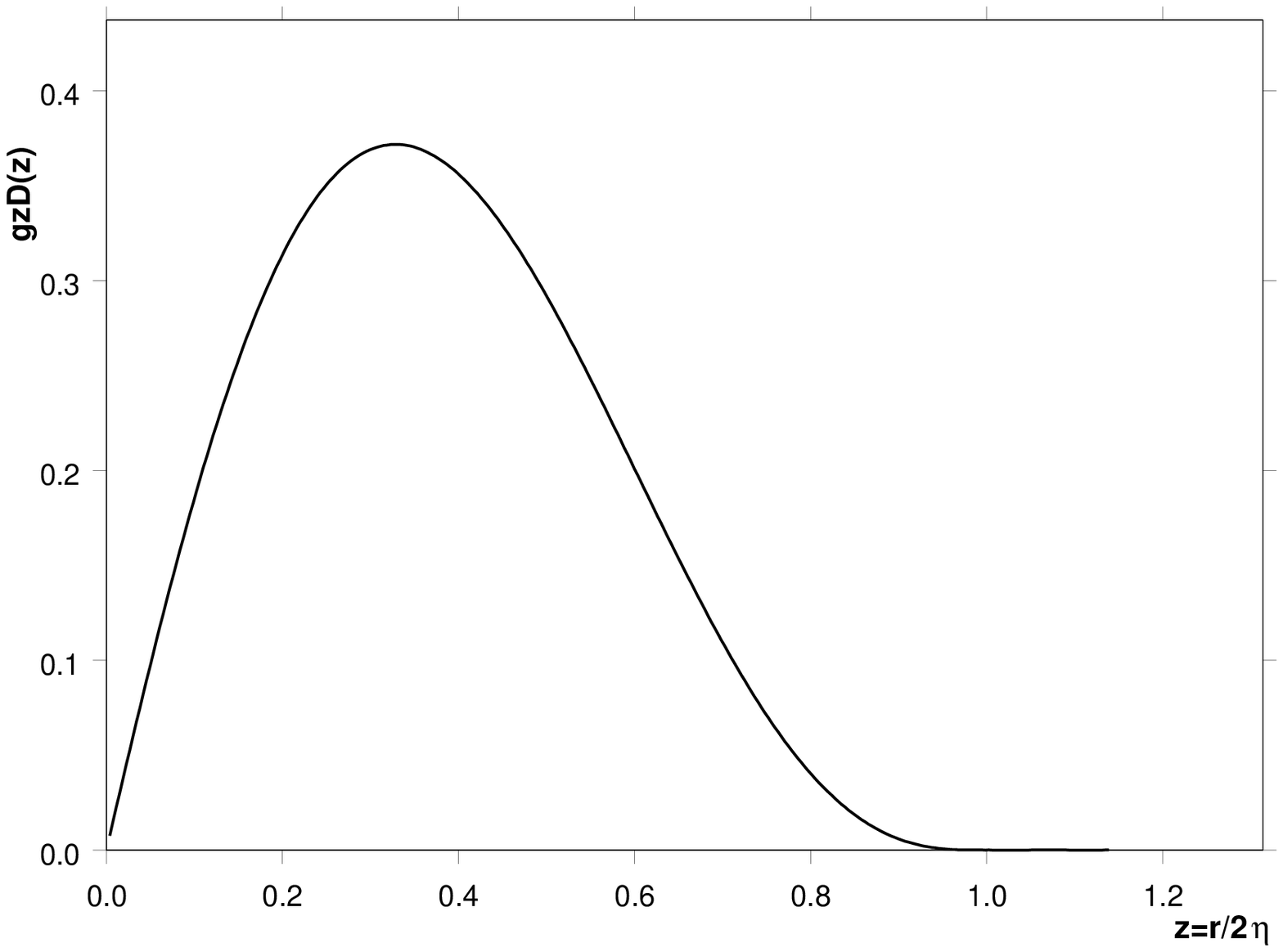,width=7in,height=8in}}
\caption{ $ gzD(z) $ vs. $ z=r/2\eta $ at $ \eta=400 $ (in units of $ m_R
$) for $ \frac{T_i}{T_c}=3 $, $ g=10^{-5} $. Radiation dominated universe.
\label{fig9} } 
\end{figure}

\begin{figure}
{\epsfig{file=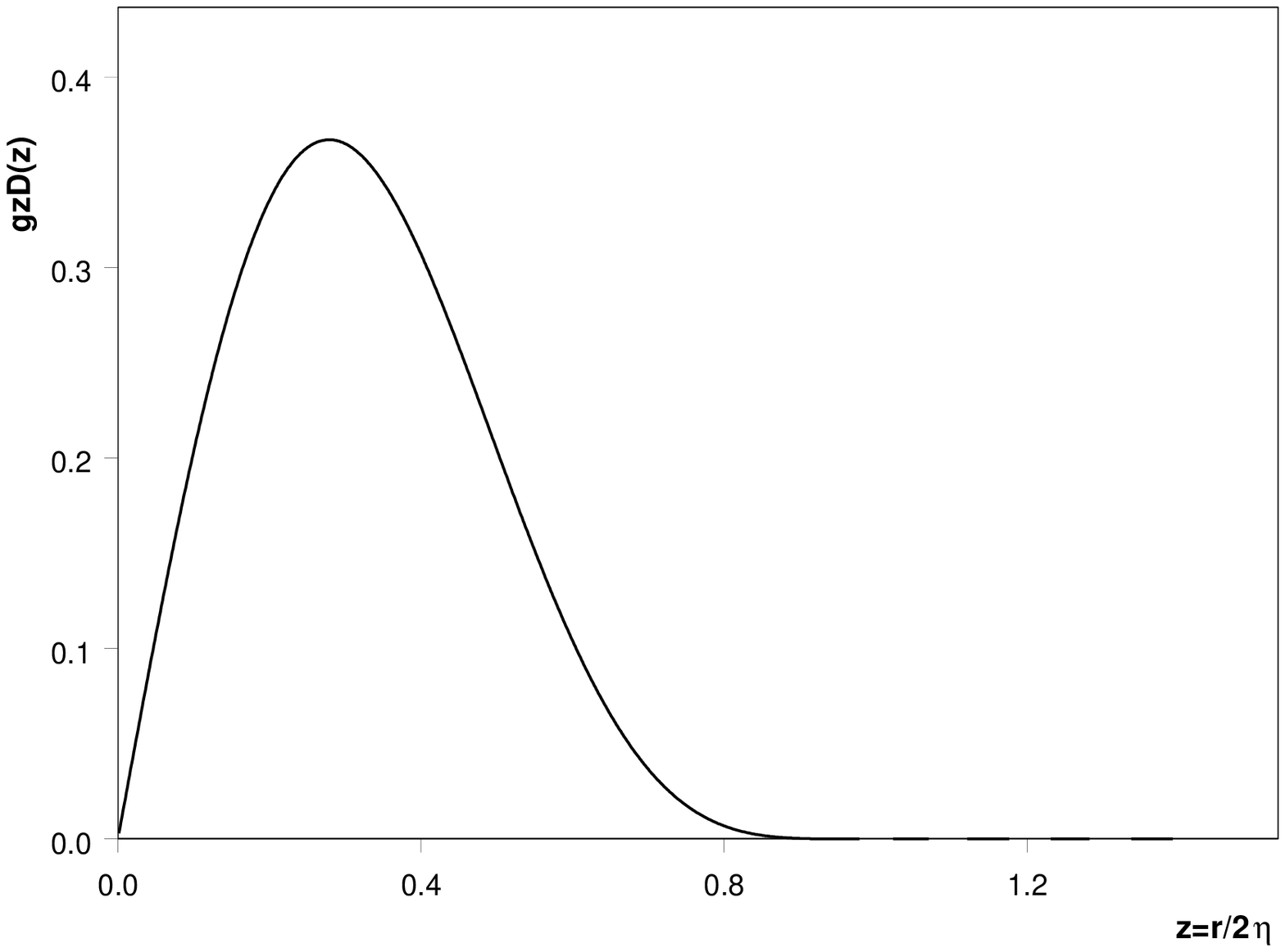,width=7in,height=8in}}
\caption{ $ gzD(z)$ vs. $ z=r/2\eta $ at $ \eta=400$  (in units of $
m_R $) for $ \frac{T_i}{T_c}=3 $, $ g=10^{-5} $. Matter dominated universe. 
\label{fig10} } 
\end{figure}

\begin{thebibliography}{99}
\bibitem{reviews}For  reviews of standard and inflationary 
cosmology see: E. W. Kolb and M. S. Turner, {\em The Early Universe}
(Addison Wesley, Redwood City, C.A. 1990). A. Linde, {\em Particle
Physics and Inflationary Cosmology}, (Harwood Academic Pub. Switzerland,
1990). R. Brandenberger, Rev. of Mod. Phys. {\bf 57},1 (1985);
Int. J. Mod. Phys. A2, 77 (1987). 

\bibitem{kibble} M. B. Hindmarsh and T.W.B. Kibble,
Rep. Prog. Phys. {\bf 58}, 477 (1995);  

A. Vilenkin and E.P.S. Shellard, `Cosmic Strings and other Topological
Defects',  Cambridge Monographs on Math. Phys. (Cambridge Univ. Press, 1994). 

\bibitem{latetimePT} C. T. Hill, D. N. Schramm and J. Fry,
Comm. Nuc. Part. Phys. {\bf 19}, 25 (1990); 

J. Friedmann, C. T. Hill and R. Watkins, Phys. Rev. {\bf D46}, 1226 (1992). 


\bibitem{boyvega}D. Boyanovsky and H. J. de Vega, Phys. Rev. {\bf
D47}, 2343 (1993); 

D. Boyanovsky, Phys. Rev.  {\bf E48}, 767 (1993);
D. Boyanovsky, D.-S. Lee and A. Singh, Phys. Rev. {\bf D48}, 800 (1993).  

\bibitem{symbreaklosala}  F. Cooper, S. Habib, Y. Kluger, E. Mottola,
 Phys.Rev. {\bf D55} (1997), 6471.

\bibitem{bosecon} D. Boyanovsky, H. J. de Vega, R. Holman and
J. Salgado, 

Phys. Rev. {\bf D59} 125009 (1999).  

\bibitem{noscorre} D. Boyanovsky, C. Destri, H.J. de Vega, R. Holman
and  J. F. J. Salgado, 

Phys.Rev. {\bf D57}, 7388 (1998).

\bibitem{nosreviews} For a detailed description of phase ordering in
quantum field theory and the similarities 
and differences with condensed matter physics see: D. Boyanovsky,
H. J. de Vega, {\em Dynamics of Symmetry Breaking Out of Equilibrium:
From Condensed Matter to QCD and the Early Universe},  hep-ph/9909372,
in the Proceedings of the VI Paris Cosmology Colloquium, June 1999,
p. , edited by H. J. de Vega and N. S\'anchez, Observatoire de Paris
publisher. To appear in the Proceedings of the National Academy of Sciences of
India, ANSI-2000. D. Boyanovsky, H. J. de Vega, R. Holman, 
 {\em Non-equilibrium phase transitions in condensed matter and
cosmology: spinodal decomposition, condensates and defects}, 
to appear in the Proceedings of the NATO ASI `Topological Defects
and the Non-Equilibrium Dynamics of Symmetry Breaking Phase
Transitions', Eds. E. Godfrin and Y. Bunkov, hep-ph/9903534.   
 
\bibitem{bray} A. J. Bray, Adv. Phys. {\bf 43}, 357 (1994);
J. S. Langer in `Solids far from Equilibrium', 
Ed. C. Godr\`eche, Cambridge Univ. Press 1992; 
J. S. Langer in `Far from Equilibrium Phase Transitions',
Ed. L. Garrido, Springer-Verlag, 1988 ; J. S. Langer 
in `Fluctuations, Instabilities and Phase Transitions', Ed. T. Riste,
Nato Advanced Study Institute, Geilo Norway, 1975, Plenum, 1975;
G. Mazenko  in `Far from Equilibrium Phase 
Transitions', Ed. L. Garrido, Springer-Verlag, 1988.  N. Goldenfeld,
`Lectures on Phase Transitions and the Renormalization Group',
Addison-Wesley, 1992.  

\bibitem{gill} For a recent review on the dynamics
of phase transitions see: A. Gill, `Contemporary Physics', vol. 39, No. 1,
p. 13, (1998). 

\bibitem{destri} C. Destri and  E. Manfredini; hep-ph/9906554. 

\bibitem{FRW} D. Boyanovsky, H. J. de Vega and R. Holman, 
Phys. Rev.  {\bf D 49}, 2769 (1994).

\bibitem{inflation} D. Boyanovsky, D. Cormier, H. J. de Vega,
R. Holman and S. P. Kumar, Phys. Rev. {\bf D 57}, 2166 (1998); D. Boyanovsky,
D. Cormier, H. J. de Vega and R. Holman, 
Phys. Rev. {\bf D55}, 3373, (1997) 3373; D. Boyanovsky, D. Cormier,
H. J. de Vega, R. Holman, A. Singh and M. Srednicki, Phys. Rev. {\bf
56}, 1939 (1997).   
 
\bibitem{nuestros} 
D. Boyanovsky, H. J. de Vega, R. Holman, D.-S. Lee and A. Singh, 
Phys. Rev. {\bf D51}, 4419 (1995). 
D. Boyanovsky, H. J. de Vega and R. Holman, Proceedings of
the Second Paris Cosmology Colloquium, Observatoire de Paris, June 1994,
pp. 127-215, H. J. de Vega and N. S\'anchez, Editors, World
Scientific, 1995; Advances in Astrofundamental Physics, Erice
Chalonge School, N. S\'anchez and A. Zichichi Editors, World
Scientific, 1995.  D. Boyanovsky, H. J. de Vega, R. Holman and J. Salgado,
Phys. Rev. {\bf D54}, 7570 (1996); 
D. Boyanovsky, H. J. de Vega and R. Holman, 
 Vth. Erice Chalonge School, Current Topics in Astrofundamental
 Physics, N. S\'anchez and A. Zichichi Editors, World Scientific,
 1996, p. 183-270. 
D. Boyanovsky, M. D'Attanasio, H. J. de Vega, R. Holman and D. S. Lee, 
Phys. Rev. {\bf D52}, 6805 (1995).

\bibitem{turok} N. Turok and D. N. Spergel, Phys. Rev. Lett. {\bf 66}, 3093
(1991); D. N. Spergel, N. Turok, W. H. Press and B. S. Ryden,
Phys. Rev. {\bf D43}, 1038 (1991); N. Turok and D. N. Spergel,
Phys. Rev. Lett. {\bf 64}, 2736 (1990).   

\bibitem{durrer}
R. Durrer, M. Kunz and A. Melchiorri
Phys. Rev. {\bf D59}, 123005 (1999), M. Kunz and R. Durrer,
Phys. Rev. {\bf D55}, R4516 (1997); R. Durrer and M. Sakellariadou,
Phys. Rev. {\bf D56}, 4480 (1997); R. Durrer and Z-H. Zhou,
Phys. Rev. {\bf D53}, 5394 (1996).   

\bibitem{kibble2} T. W. B. Kibble, J. Phys. A 9, 1387 (1976) 

\bibitem{losalamos}  F. Cooper, S. Habib, Y. Kluger,
E. Mottola, J. P. Paz, P. R. Anderson, 
Phys. Rev. {\bf D50}, 2848 (1994). 
F. Cooper, Y. Kluger, E. Mottola, J. P. Paz, Phys. Rev. {\bf D51},
2377 (1995); F. Cooper and E. Mottola, Mod. Phys. Lett. {\bf A 2}, 635 (1987);
F. Cooper and E. Mottola, Phys. Rev. {\bf D36}, 3114 (1987);
F. Cooper, S.-Y. Pi and P. N. Stancioff, Phys. Rev. {\bf D34}, 3831 (1986). 

\bibitem{baacke} J. Baacke, K. Heitmann and C. Patzold, Phys.Rev. {\bf
D56} (1997) 6556;  Phys.Rev.{\bf D55}, (1997) 2320, J. Baacke and
C. Patzold,  hep-ph/9906417. 

\bibitem{lebellac} M. Le Bellac, {\em Thermal Field Theory}, Cambridge
University Press, Cambridge, England, 1996.  

\bibitem{parwani} R.R. Parwani, Phys. Rev. {\bf D45}, 4695 (1992).

\bibitem{polonyi} J. Alexandre and J. Polonyi,  hep-th/9906017;
J. Alexandre, V. Branchina and  J. Polonyi, Phys. Rev. {\bf  D58} 016002
(1998); J. Alexandre, PhD. Thesis (unpublished); J. Alexandre,
V. Branchina and J. Polonyi, Phys. Lett. {\bf B445}, 351 (1999).  

\bibitem{wetterich} A. Ringwald and C. Wetterich, Nucl. Phys. {\bf B
334}, 506 (1990), N. Tetradis and C. Wetterich, Nucl. Phys. {\bf B
383}, 197 (1992); N. Tetradis, Nucl. Phys. {\bf B 488}, 92 (1997).   

\bibitem{gr} I. S. Gradshteyn and I. M. Ryshik, Table of Integrals, Series and
Products, Academic Press. 
\bibitem{prud} A. P. Prudnikov, Yu. A. Brichkov and O. I. Marichev,
Integrals and series, vol III, Nauka, Moscow, 1981.
\end{thebibliography}
\end{document}